\documentclass[prd,twocolumn,showpacs,superscriptaddress,floatfix]{revtex4}

\usepackage{slashed}
\usepackage{epsfig}
\usepackage{amsmath}
\usepackage{comment}
\usepackage{longtable}
\usepackage{subfigure}
\usepackage{color}
\usepackage{rotating}

\def\met{{\mbox{$E\kern-0.57em\raise0.19ex\hbox{/}_{T}$}}}

\def\metbb{$WH,ZH\rightarrow\met b\bar{b}$}
\def\WH{$WH\rightarrow \ell\nu b\bar{b}$}

\def\lmet{$WH\rightarrow \ell\kern-0.45em\raise0.19ex\hbox{/} \nu b\bar{b}$}
\def\ZH{$ZH\rightarrow \ell^+\ell^- b\bar{b}$}

\def\hww{$H\rightarrow W^+ W^-$}

\def\hzz{$H\rightarrow ZZ$}
\def\hgg{$H\rightarrow \gamma \gamma$}
\def\hbb{$H\rightarrow b\bar{b}$}
\def\Zbb{$Z\rightarrow b\bar{b}$}

\def\gevcc{GeV/$c^2$}
\newcommand{\METVEC}{\mbox{$\raisebox{.3ex}{$\not\!$}{\vec E}_T$}}

\newcommand{\PYTHIA}     {{\sc pythia}}

\newcommand{\MET}{$\not\!\!E_T$}

\begin{document}

\title{Combination of searches for the Higgs boson using the full CDF data set}

\affiliation{Institute of Physics, Academia Sinica, Taipei, Taiwan 11529, Republic of China}
\affiliation{Argonne National Laboratory, Argonne, Illinois 60439, USA}
\affiliation{University of Athens, 157 71 Athens, Greece}
\affiliation{Institut de Fisica d'Altes Energies, ICREA, Universitat Autonoma de Barcelona, E-08193, Bellaterra (Barcelona), Spain}
\affiliation{Baylor University, Waco, Texas 76798, USA}
\affiliation{Istituto Nazionale di Fisica Nucleare Bologna, $^{ee}$University of Bologna, I-40127 Bologna, Italy}
\affiliation{University of California, Davis, Davis, California 95616, USA}
\affiliation{University of California, Los Angeles, Los Angeles, California 90024, USA}
\affiliation{Instituto de Fisica de Cantabria, CSIC-University of Cantabria, 39005 Santander, Spain}
\affiliation{Carnegie Mellon University, Pittsburgh, Pennsylvania 15213, USA}
\affiliation{Enrico Fermi Institute, University of Chicago, Chicago, Illinois 60637, USA}
\affiliation{Comenius University, 842 48 Bratislava, Slovakia; Institute of Experimental Physics, 040 01 Kosice, Slovakia}
\affiliation{Joint Institute for Nuclear Research, RU-141980 Dubna, Russia}
\affiliation{Duke University, Durham, North Carolina 27708, USA}
\affiliation{Fermi National Accelerator Laboratory, Batavia, Illinois 60510, USA}
\affiliation{University of Florida, Gainesville, Florida 32611, USA}
\affiliation{Laboratori Nazionali di Frascati, Istituto Nazionale di Fisica Nucleare, I-00044 Frascati, Italy}
\affiliation{University of Geneva, CH-1211 Geneva 4, Switzerland}
\affiliation{Glasgow University, Glasgow G12 8QQ, United Kingdom}
\affiliation{Harvard University, Cambridge, Massachusetts 02138, USA}
\affiliation{Division of High Energy Physics, Department of Physics, University of Helsinki and Helsinki Institute of Physics, FIN-00014, Helsinki, Finland}
\affiliation{University of Illinois, Urbana, Illinois 61801, USA}
\affiliation{The Johns Hopkins University, Baltimore, Maryland 21218, USA}
\affiliation{Institut f\"{u}r Experimentelle Kernphysik, Karlsruhe Institute of Technology, D-76131 Karlsruhe, Germany}
\affiliation{Center for High Energy Physics: Kyungpook National University, Daegu 702-701, Korea; Seoul National University, Seoul 151-742, Korea; Sungkyunkwan University, Suwon 440-746, Korea; Korea Institute of Science and Technology Information, Daejeon 305-806, Korea; Chonnam National University, Gwangju 500-757, Korea; Chonbuk National University, Jeonju 561-756, Korea; Ewha Womans University, Seoul, 120-750, Korea}
\affiliation{Ernest Orlando Lawrence Berkeley National Laboratory, Berkeley, California 94720, USA}
\affiliation{University of Liverpool, Liverpool L69 7ZE, United Kingdom}
\affiliation{University College London, London WC1E 6BT, United Kingdom}
\affiliation{Centro de Investigaciones Energeticas Medioambientales y Tecnologicas, E-28040 Madrid, Spain}
\affiliation{Massachusetts Institute of Technology, Cambridge, Massachusetts 02139, USA}
\affiliation{Institute of Particle Physics: McGill University, Montr\'{e}al, Qu\'{e}bec H3A~2T8, Canada; Simon Fraser University, Burnaby, British Columbia V5A~1S6, Canada; University of Toronto, Toronto, Ontario M5S~1A7, Canada; and TRIUMF, Vancouver, British Columbia V6T~2A3, Canada}
\affiliation{University of Michigan, Ann Arbor, Michigan 48109, USA}
\affiliation{Michigan State University, East Lansing, Michigan 48824, USA}
\affiliation{Institution for Theoretical and Experimental Physics, ITEP, Moscow 117259, Russia}
\affiliation{University of New Mexico, Albuquerque, New Mexico 87131, USA}
\affiliation{The Ohio State University, Columbus, Ohio 43210, USA}
\affiliation{Okayama University, Okayama 700-8530, Japan}
\affiliation{Osaka City University, Osaka 588, Japan}
\affiliation{University of Oxford, Oxford OX1 3RH, United Kingdom}
\affiliation{Istituto Nazionale di Fisica Nucleare, Sezione di Padova-Trento, $^{ff}$University of Padova, I-35131 Padova, Italy}
\affiliation{University of Pennsylvania, Philadelphia, Pennsylvania 19104, USA}
\affiliation{Istituto Nazionale di Fisica Nucleare Pisa, $^{gg}$University of Pisa, $^{hh}$University of Siena and $^{ii}$Scuola Normale Superiore, I-56127 Pisa, Italy, $^{mm}$INFN Pavia and University of Pavia, I-27100 Pavia, Italy}
\affiliation{University of Pittsburgh, Pittsburgh, Pennsylvania 15260, USA}
\affiliation{Purdue University, West Lafayette, Indiana 47907, USA}
\affiliation{University of Rochester, Rochester, New York 14627, USA}
\affiliation{The Rockefeller University, New York, New York 10065, USA}
\affiliation{Istituto Nazionale di Fisica Nucleare, Sezione di Roma 1, $^{jj}$Sapienza Universit\`{a} di Roma, I-00185 Roma, Italy}
\affiliation{Texas A\&M University, College Station, Texas 77843, USA}
\affiliation{Istituto Nazionale di Fisica Nucleare Trieste/Udine; $^{nn}$University of Trieste, I-34127 Trieste, Italy; $^{kk}$University of Udine, I-33100 Udine, Italy}
\affiliation{University of Tsukuba, Tsukuba, Ibaraki 305, Japan}
\affiliation{Tufts University, Medford, Massachusetts 02155, USA}
\affiliation{University of Virginia, Charlottesville, Virginia 22906, USA}
\affiliation{Waseda University, Tokyo 169, Japan}
\affiliation{Wayne State University, Detroit, Michigan 48201, USA}
\affiliation{University of Wisconsin, Madison, Wisconsin 53706, USA}
\affiliation{Yale University, New Haven, Connecticut 06520, USA}

\author{T.~Aaltonen}
\affiliation{Division of High Energy Physics, Department of Physics, University of Helsinki and Helsinki Institute of Physics, FIN-00014, Helsinki, Finland}
\author{S.~Amerio}
\affiliation{Istituto Nazionale di Fisica Nucleare, Sezione di Padova-Trento, $^{ff}$University of Padova, I-35131 Padova, Italy}
\author{D.~Amidei}
\affiliation{University of Michigan, Ann Arbor, Michigan 48109, USA}
\author{A.~Anastassov$^x$}
\affiliation{Fermi National Accelerator Laboratory, Batavia, Illinois 60510, USA}
\author{A.~Annovi}
\affiliation{Laboratori Nazionali di Frascati, Istituto Nazionale di Fisica Nucleare, I-00044 Frascati, Italy}
\author{J.~Antos}
\affiliation{Comenius University, 842 48 Bratislava, Slovakia; Institute of Experimental Physics, 040 01 Kosice, Slovakia}
\author{G.~Apollinari}
\affiliation{Fermi National Accelerator Laboratory, Batavia, Illinois 60510, USA}
\author{J.A.~Appel}
\affiliation{Fermi National Accelerator Laboratory, Batavia, Illinois 60510, USA}
\author{T.~Arisawa}
\affiliation{Waseda University, Tokyo 169, Japan}
\author{A.~Artikov}
\affiliation{Joint Institute for Nuclear Research, RU-141980 Dubna, Russia}
% ...

\collaboration{CDF Collaboration\footnote{With visitors from
$^a$Istituto Nazionale di Fisica Nucleare, Sezione di Cagliari, 09042 Monserrato (Cagliari), Italy,
$^b$University of California Irvine, Irvine, CA 92697, USA,
$^c$University of California Santa Barbara, Santa Barbara, CA 93106, USA,
$^d$University of California Santa Cruz, Santa Cruz, CA 95064, USA,
$^e$Institute of Physics, Academy of Sciences of the Czech Republic, 182~21, Czech Republic,
$^f$CERN, CH-1211 Geneva, Switzerland,
$^g$Cornell University, Ithaca, NY 14853, USA,
$^h$University of Cyprus, Nicosia CY-1678, Cyprus,
$^i$Office of Science, U.S. Department of Energy, Washington, DC 20585, USA,
$^j$University College Dublin, Dublin 4, Ireland,
$^k$ETH, 8092 Z\"{u}rich, Switzerland,
$^l$University of Fukui, Fukui City, Fukui Prefecture, Japan 910-0017,
$^m$Universidad Iberoamericana, Lomas de Santa Fe, M\'{e}xico, C.P. 01219, Distrito Federal,
$^n$University of Iowa, Iowa City, IA 52242, USA,
$^o$Kinki University, Higashi-Osaka City, Japan 577-8502,
$^p$Kansas State University, Manhattan, KS 66506, USA,
$^q$Brookhaven National Laboratory, Upton, NY 11973, USA,
$^r$University of Manchester, Manchester M13 9PL, United Kingdom,
$^s$Queen Mary, University of London, London, E1 4NS, United Kingdom,
$^t$University of Melbourne, Victoria 3010, Australia,
$^u$Muons, Inc., Batavia, IL 60510, USA,
$^v$Nagasaki Institute of Applied Science, Nagasaki 851-0193, Japan,
$^w$National Research Nuclear University, Moscow 115409, Russia,
$^x$Northwestern University, Evanston, IL 60208, USA,
$^y$University of Notre Dame, Notre Dame, IN 46556, USA,
$^z$Universidad de Oviedo, E-33007 Oviedo, Spain,
$^{aa}$CNRS-IN2P3, Paris, F-75205 France,
$^{bb}$Texas Tech University, Lubbock, TX 79609, USA,
$^{cc}$Universidad Tecnica Federico Santa Maria, 110v Valparaiso, Chile,
$^{dd}$Yarmouk University, Irbid 211-63, Jordan,
$^{ll}$Universite catholique de Louvain, 1348 Louvain-La-Neuve, Belgium,
$^{oo}$University of Z\"{u}rich, 8006 Z\"{u}rich, Switzerland,
$^{pp}$Massachusetts General Hospital and Harvard Medical School, Boston, MA 02114 USA,
$^{qq}$Hampton University, Hampton, VA 23668, USA,
$^{rr}$Los Alamos National Laboratory, Los Alamos, NM 87544, USA
}}
\noaffiliation

\date{\today}

\begin{abstract}
We present a combination of searches for the standard model Higgs
boson using the full CDF Run II data set, which corresponds to 
an integrated luminosity of 9.45--10.0 fb$^{-1}$ collected from
$\sqrt{s}=1.96$~TeV $p{\bar{p}}$ collisions at the Fermilab Tevatron.
The searches consider Higgs boson production from gluon-gluon fusion,
vector-boson fusion, and associated production with either a $W$ or
$Z$ boson or a $t{\bar{t}}$ pair.  Depending on the production mode,
Higgs boson decays to $W^+W^-$, $ZZ$, $b{\bar{b}}$, $\tau^+\tau^-$,
and $\gamma\gamma$ are examined.  We search for a Higgs boson with
masses ($m_H$) in the range 90--200~GeV/$c^2$.  In the absence of 
a signal, we expect based on combined search sensitivity to exclude 
at the 95\% credibility level the mass regions $90<m_H<94$~GeV/$c^2$, 
$96<m_H<106$~GeV/$c^2$, and $153<m_H<175$~GeV/$c^2$.  The observed 
exclusion regions are $90<m_H<102$~GeV/$c^2$ and $149<m_H<172$~GeV/$c^2$.    
A moderate excess of signal-like events relative to the background
expectation at the level of 2.0 standard deviations is present in 
the data for the $m_H=125$~GeV/$c^2$ search hypothesis.  We also 
present interpretations of the data within the context of a 
fermiophobic model and an alternative standard model incorporating 
a fourth generation of fermions. Finally, for the hypothesis of a 
new particle with mass $125$~GeV/$c^2$, we constrain the coupling
strengths of the new particle to $W^\pm$ bosons, $Z$ bosons, and
fermions.
\end{abstract}

\pacs{13.85.Rm, 14.80.Bn}

\maketitle

\section{Introduction}
\label{sec:intro}

Within the standard model (SM)~\cite{gws} of particle physics, the 
mechanism of electroweak symmetry breaking~\cite{higgs} implies the 
existence of a single observable particle referred to as the Higgs 
boson, $H$.  The mass of this neutral scalar is not predicted by the 
theoretical framework of the SM and must be measured experimentally.
Similarly, Yukawa couplings between fundamental fermions and the Higgs 
field, which are responsible for fermion masses, are not predicted by 
the SM.

Precision electroweak measurements from LEP, SLC, and the Tevatron
have been interpreted within the context of the SM to constrain the
mass of the potential SM Higgs boson~\cite{elweak}.  Including the
most recent $W$ boson and top-quark mass measurements from the
Tevatron~\cite{wmass,topmass}, the electroweak data are consistent 
with a Higgs boson mass smaller than 152 GeV/$c^2$ at the 95\% 
confidence level, within the framework of the SM.  Direct searches 
at LEP exclude the SM Higgs boson for masses less than 114.4
GeV/$c^2$~\cite{sm-lep}.

Recently, a new particle was observed in data collected from
$\sqrt{s}=$7--8~TeV $pp$ collisions at the CERN Large Hadron Collider
(LHC) by the ATLAS~\cite{atlasdiscovery} and CMS~\cite{cmsdiscovery}
collaborations.  The reported measurements of the observed particle
are consistent with the expectations for the SM Higgs boson with a
mass of roughly 125 GeV/$c^2$.  The specific final states contributing
the greatest amount of significance to these observations are 
$\gamma\gamma$ and $ZZ\rightarrow\ell\ell\ell\ell$~\cite{notation}.  
Complementary evidence was also reported recently in the $b \bar{b}$ 
final state based on a combination of searches from the CDF and D0 
experiments~\cite{tevbbevidence}.  Precision measurements of the 
properties of the new particle such as its spin, parity, production 
rates via the different mechanisms, and decay branching ratios are 
necessary for identifying if the new particle is in fact the SM Higgs 
boson.  Higgs boson searches at the Tevatron obtain most of their 
sensitivity from production and decay modes that are different from 
those of the LHC searches.  Tevatron measurements therefore provide 
important contributions to the available constraints on several of 
these properties.
 
The SM Higgs boson production process with the largest cross section 
at the Tevatron is gluon fusion. Associated production with a $W$ 
or $Z$ boson ($VH$) is the second largest.  The cross section for 
{\it WH} production is twice that of {\it ZH} and is about a factor 
of ten smaller than gluon fusion. The Higgs boson decay branching 
fraction is dominated by \hbb\ for the {\it low-mass} Higgs boson 
($m_H < 135$~\gevcc) and by \hww\ for the {\it high-mass} Higgs boson 
($m_H >135$ \gevcc). An inclusive search for the low-mass Higgs boson 
in the \hbb\ decay channel is extremely challenging because the 
$b\bar{b}$ production rate through SM processes is many orders of 
magnitude larger than that expected from the Higgs boson production 
rate.  Requiring the leptonic decay of the associated $W$ or $Z$ boson 
greatly improves the expected signal-to-background ratio in these 
channels.  As a result, Higgs associated production followed by 
the \hbb\ decay is the most promising channel in searches for the 
low-mass Higgs boson.  For higher-mass Higgs boson searches, the 
\hww\ decay, where leptons originate from the $W$ boson decays, are 
the most sensitive.  While the \hbb\ and \hww\ search channels provide 
the best sensitivity, searches made in all final states are combined 
to obtain the highest possible sensitivity to the SM Higgs boson.

This article presents a combination of CDF searches for the SM Higgs
boson.  The combined searches incorporate potential contributions from
Higgs boson production via gluon fusion, production in association
with a $W$ or $Z$ boson, vector-boson fusion (VBF) production, and
production in association with a top-quark pair.  Higgs boson decay
modes considered are $H\rightarrow W^+W^-$, $H\rightarrow ZZ$, \hbb,
$H\rightarrow \tau^+\tau^-$, and $H\rightarrow \gamma\gamma$.  The
individual searches are performed for potential Higgs boson masses in
the range from 90 to 200~GeV/$c^2$ using non-overlapping data sets
defined by distinct final states.  For each search sub-channel, SM
backgrounds are estimated and validated using data events populating
appropriately-defined control regions.  Finally, a discriminant, 
which is typically the output of a multivariate algorithm constructed 
from kinematic event variables, is used to separate a potential signal 
from much larger background event contributions.  The multivariate
algorithms are separately optimized for each Higgs boson mass
hypothesis and for each analysis sub-channel.  Search results are
combined by constructing a combined likelihood function based on the
final discriminant distributions in each search sub-channel, taking
into account the correlations among channels.  After performing the
combined search over the full Higgs boson mass range, we focus on 
the 125 GeV/$c^2$ mass hypothesis, motivated by the recent ATLAS and 
CMS observations~\cite{atlasdiscovery,cmsdiscovery}.  Assuming the 
LHC signal is present in CDF data, we constrain fermion and boson 
couplings to this new particle.
 
We additionally interpret the search results within the context of a fermiophobic 
Higgs model (FHM)~\cite{fermiophobic1,fermiophobic2,fermiophobic3,fermiophobic4},
which assumes SM couplings to the Higgs boson except in the case of fermions, 
for which the couplings are assumed to vanish.  In this model, gluon-fusion 
production is highly suppressed, while branching fractions for $H\rightarrow
\gamma\gamma$, $H\rightarrow W^+W^-$, and $H\rightarrow ZZ$ are enhanced.  We 
also consider an extension of the SM incorporating a heavy fourth generation 
of fermions (SM4). Within this model, gluon-fusion production is significantly 
enhanced~\cite{sm41,sm42,sm43}.

This article is organized as follows: Section~\ref{sec:detector} briefly
describes the CDF~II detector and the data samples used for this
combination; Section~\ref{sec:signal} describes the predictions for
Higgs boson production and decay that are assumed throughout, as well as 
the Monte Carlo models used to predict the differential distributions;
Section~\ref{sec:channels} describes the search channels included in
the combination; Section~\ref{sec:systematics} describes the dominant
sources of uncertainty in each channel and the correlations of these 
uncertainties between channels; Section~\ref{sec:statistics} describes 
the statistical methods used; Section~\ref{sec:smresults} presents results 
in the context of the SM; Section~\ref{sec:fp} presents results in the 
context of the fermiophobic model; Section~\ref{sec:sm4} presents results 
in the context of the SM4 model; Section ~\ref{sec:couplings} describes 
the measurement of fermion and boson couplings in the context of a new 
125 GeV/$c^2$ boson; and Section~\ref{sec:summary} summarizes the article.

\section{The CDF II detector and the full CDF data set} 
\label{sec:detector}

The CDF~II detector is described in detail elsewhere~\cite{secvtx,cdfdetector}.  
Silicon-strip tracking detectors~\cite{cdfsilicon} surround the interaction 
region and provide precision measurements of charged-particle trajectories in 
the range $|\eta|<2$~\cite{coord}.  A cylindrical drift chamber provides full 
coverage over the range $|\eta|<1$.  The tracking detectors are located within 
a 1.4~T superconducting solenoidal magnet with field oriented along the beam 
direction.  The energies of individual particles and particle jets are 
measured in segmented electromagnetic and hadronic calorimeter modules 
arranged in a projective tower geometry surrounding the solenoid.  Ionization 
chambers are located outside of the calorimeters to help identify muon 
candidates~\cite{cdfmuons}.  The Tevatron collider luminosity is measured 
with multi-cell gas Cherenkov detectors~\cite{CLC}.  The total uncertainty on 
luminosity measurements is $\pm 6.0$\%, of which 4.4\% originates from detector 
acceptance uncertainties and 4.0\% is due to the uncertainty on the inelastic 
$p{\bar{p}}$ cross section~\cite{inelppbarxs}.

All of the results combined in this article, with the exception of the
$H\rightarrow\tau^+\tau^-$ search, use the full CDF Run~II data sample.  
Small variations in the luminosities reported for the different search 
channels reflect the application of channel-specific data-quality criteria 
designed to ensure proper data modeling.  For example, the silicon detector 
is required to be operational for the $H\rightarrow b{\bar{b}}$ searches, 
for which the identification of secondary track vertices from $b$ hadrons 
plays an important role, but not in the case of the $H\rightarrow\gamma
\gamma$ search.  The $H\rightarrow\gamma\gamma$ search makes use of the 
largest data set, 10~fb$^{-1}$, which is about 82~\% of the 12~fb$^{-1}$ 
that was delivered by the Tevatron collider and about 99.5~\% of the 
luminosity in which the CDF detector was considered to be operational.

CDF uses a three-level online event selection system (trigger) to select 
beam collision events at a rate that can be written into permanent storage.  
The first trigger level relies on special-purpose hardware~\cite{XFT} to 
reduce the effective beam-crossing frequency of 1.7~MHz to an event rate 
of approximately 15~kHz.  The second level uses a mixture of dedicated
hardware and fast software algorithms to further reduce the event rate 
to roughly 1~kHz.  Events satisfying level-two trigger requirements are 
read out of the detector and passed to an array of computers running fast 
versions of offline reconstruction algorithms, which allow for third-level 
trigger decisions based on quantities that are nearly the same as those 
used in offline analyses~\cite{cdfl3}.  The final rate of events written 
into permanent storage is approximately 100~Hz.  The basic trigger criteria 
for events used in these searches are the presence of high-transverse 
momentum ($p_T$)~\cite{coord} leptons, clustered calorimeter energy 
deposits associated with partons originating from the collision (jets)~\cite{JES}, 
and large imbalances in the transverse energies ($E_T$)~\cite{coord} of measured 
depositions within the calorimeter, associated with evidence for undetected 
neutrinos within the event (\met)~\cite{coord}. 

\section{Standard model Higgs boson signal predictions}
\label{sec:signal}

In order to conduct the most sensitive Higgs boson search possible, 
we include contributions from all significant production modes that 
are expected to occur at the Tevatron.  When conducting the search
using multiple production modes, the predictions of the relative
contributions of each mode and the uncertainties on those predictions
are required.  In addition, because we use multivariate analysis
techniques, the predictions of the kinematic distributions for the
signal model are also important.  Here we provide a summary of the
tools we use to make predictions for the Higgs boson signal.  The
theoretical uncertainties on the signal model play a significant 
role at higher masses where gluon fusion is the major production 
mode, but are less important for low-mass Higgs boson searches where 
associated production is most important.

To predict the kinematic distributions of Higgs boson signal events,
we use the \textsc{pythia}~\cite{pythia} Monte Carlo program, with
CTEQ5L~\cite{cteq} leading-order (LO) parton distribution functions 
(PDFs).  We scale these Monte Carlo predictions to the highest-order 
cross section calculations available.  The \textsc{pythia} differential
distributions for some important variables, such as the Higgs boson
$p_T$ and the number of associated jets, are also corrected based on
higher-order calculations as described below.  The $gg\rightarrow H$
production cross section is calculated at next-to-next-to leading order 
(NNLO) in quantum chromodynamics (QCD) with a next-to-next-to leading 
log (NNLL) resummation of soft gluons; the calculation also includes 
two-loop electroweak effects and handling of the running $b$-quark 
mass~\cite{anastasiou,grazzinideflorian}.  The numerical values in 
Table~\ref{tab:higgsxsec} are updates~\cite{grazziniprivate} of 
the predictions in~\cite{grazzinideflorian} with $m_t$ set to 
173.1~GeV/$c^2$~\cite{topmass}, and with a treatment of the massive 
top and bottom loop corrections up to next-to-leading-order (NLO) 
and next-to-leading-log (NLL) accuracy. For these calculations the 
factorization scale ($\mu_F$) and renormalization scale ($\mu_R$) 
are set to $\mu_F=\mu_R=m_H$, and the MSTW~2008 NNLO PDF 
set~\cite{mstw2008}, as recommended by the PDF4LHC working 
group~\cite{pdf4lhc}, is used.  The calculations are refinements of 
earlier NNLO calculations of the $gg\rightarrow H$ production cross
section~\cite{harlanderkilgore2002,anastasioumelnikov2002,ravindran2003}.
Electroweak corrections were computed in Refs.~\cite{actis2008,aglietti}. 
Soft gluon resummation was introduced in the prediction of the 
$gg\rightarrow H$ production cross section in Ref.~\cite{catani2003}.  
The $gg\rightarrow H$ production cross section depends strongly on the 
gluon PDF and the value of the strong interaction coupling constant
corresponding to the value $q$ of transferred momentum, $\alpha_s(q^2)$.

Analyses consider $gg\rightarrow H$ production are either
treated inclusively, or are divided into categories based on the
number of reconstructed jets.  This division is described in
Table~\ref{tab:cdfacc}.  For analyses that consider inclusive
$gg\rightarrow H$ production we use uncertainties calculated from
simultaneous variation of the factorization and renormalization scales
by factors of two.  We use the prescription of the PDF4LHC working
group~\cite{pdf4lhc} for evaluating PDF uncertainties on the inclusive
production cross section.  QCD scale uncertainties that affect the
cross section through their impacts on the PDFs are included as a
correlated part of the total scale uncertainty.  The remainder of the
PDF uncertainty is treated as uncorrelated with the QCD scale
uncertainty.

For analyses seeking $gg\rightarrow H$ production that divide events
into categories based on the number of reconstructed jets(see Table~\ref{tab:cdfacc}), we follow
Refs.~\cite{bnlaccord,lhcdifferential} for evaluating the impacts of
the scale uncertainties.  We treat the QCD scale uncertainties
obtained from the NNLL inclusive~\cite{grazzinideflorian,anastasiou},
NLO one or more jets~\cite{anastasiouwebber}, and NLO two or more
jets~\cite{campbellh2j} cross section calculations as uncorrelated
with one another.  We then obtain QCD scale uncertainties for the
exclusive $gg\rightarrow H+0$~jet, 1~jet, and 2~or more jet categories
by propagating the uncertainties on the inclusive cross section
predictions through the subtractions needed to predict the exclusive
rates.  For example, the $H$+0~jet cross section is obtained by
subtracting the cross section for production of Higgs bosons with 
one or more jets at NLO from the inclusive NNLL+NNLO cross section.  
We assign three separate, uncorrelated scale uncertainties with 
correlated and anticorrelated contributions among exclusive jet 
categories.  The procedure in Ref.~\cite{anastasiouwebber} is used 
to determine PDF model uncertainties.  These are obtained separately 
for each bin of jet multiplicity and treated as 100\% correlated 
among jet bins.

The scale choice affects the $p_T$ spectrum of the Higgs boson when
produced in gluon-gluon fusion, thus biasing the acceptance
of the selection requirements and also the shapes of the distributions
of the final discriminants.  The effect of the acceptance change is
included in the calculations of Ref.~\cite{anastasiouwebber} and
Ref.~\cite{campbellh2j}, as the experimental requirements are simulated
in these calculations. The effects on the final discriminant shapes
are obtained by reweighting the $p_T$ spectrum of the Higgs boson
production in the Monte Carlo simulations to higher-order calculations.
The Monte Carlo signal simulation used is provided by
the LO generator {\sc pythia}~\cite{pythia}, which includes a parton
shower and fragmentation and hadronization models.   We reweight the
Higgs boson $p_T$ spectra in the {\sc pythia} Monte Carlo samples to
that predicted by {\sc hqt}~\cite{hqt} when making predictions of
differential distributions of $gg\rightarrow H$ signal events. To
evaluate the impact of the scale uncertainty on the differential
spectra, we use the {\sc resbos}~\cite{resbos} generator, apply
the scale-dependent differences in the Higgs boson $p_T$ spectrum to
the {\sc hqt} prediction, and propagate these to the final
discriminants as a systematic uncertainty on the shape, which is
included in the calculation of the limits.

\begin{sidewaystable*}{}
\begin{center}
\caption{
The production cross sections and decay branching fractions for the SM
Higgs boson assumed for the combination.}
\vspace{0.2cm}
\label{tab:higgsxsec}
\begin{ruledtabular}
\begin{tabular}{ccccccccccc}
$m_H$ & $\sigma_{gg\rightarrow H}$ & $\sigma_{{\it WH}}$ & $\sigma_{{\it ZH}}$ & $\sigma_{{\rm VBF}}$ & $\sigma_{ttH}$  &
$Br(H\rightarrow b{\bar{b}})$ & $Br(H\rightarrow \tau^+\tau^-)$ & $Br(H\rightarrow W^+W^-)$ & $Br(H\rightarrow ZZ)$ & $Br(H\rightarrow\gamma\gamma)$ \\
(GeV/$c^2$) & (fb)  & (fb)    & (fb)    & (fb)   & (fb)   & (\%)     & (\%)    & (\%)   & (\%)  & (\%)    \\ \hline
\hline
    90 &  2442.3    &  394.7  & 224.0   & 114.8  &        & 81.2     & 8.41    & 0.21   & 0.04  & 0.123   \\
    95 &  2101.1    &  332.1  & 190.3   & 105.6  &        & 80.4     & 8.41    & 0.47   & 0.07  & 0.140   \\
   100 &  1821.8    &  281.1  & 162.7   &  97.3  &  8.0   & 79.1     & 8.36    & 1.11   & 0.11  & 0.159   \\
   105 &  1584.7    &  238.7  & 139.5   &  89.8  &  7.1   & 77.3     & 8.25    & 2.43   & 0.22  & 0.178   \\
   110 &  1385.0    &  203.7  & 120.2   &  82.8  &  6.2   & 74.5     & 8.03    & 4.82   & 0.44  & 0.197   \\
   115 &  1215.9    &  174.5  & 103.9   &  76.5  &  5.5   & 70.5     & 7.65    & 8.67   & 0.87  & 0.213   \\
   120 &  1072.3    &  150.1  &  90.2   &  70.7  &  4.9   & 64.9     & 7.11    & 14.3   & 1.60  & 0.225   \\
   125 &   949.3    &  129.5  &  78.5   &  65.3  &  4.3   & 57.8     & 6.37    & 21.6   & 2.67  & 0.230   \\
   130 &   842.9    &  112.0  &  68.5   &  60.5  &  3.8   & 49.4     & 5.49    & 30.5   & 4.02  & 0.226   \\
   135 &   750.8    &   97.2  &  60.0   &  56.0  &  3.3   & 40.4     & 4.52    & 40.3   & 5.51  & 0.214   \\
   140 &   670.6    &   84.6  &  52.7   &  51.9  &  2.9   & 31.4     & 3.54    & 50.4   & 6.92  & 0.194   \\
   145 &   600.6    &   73.7  &  46.3   &  48.0  &  2.6   & 23.1     & 2.62    & 60.3   & 7.96  & 0.168   \\
   150 &   539.1    &   64.4  &  40.8   &  44.5  &  2.3   & 15.7     & 1.79    & 69.9   & 8.28  & 0.137   \\
   155 &   484.0    &   56.2  &  35.9   &  41.3  &  2.0   & 9.18     & 1.06    & 79.6   & 7.36  & 0.100   \\
   160 &   432.3    &   48.5  &  31.4   &  38.2  &  1.8   & 3.44     & 0.40    & 90.9   & 4.16  & 0.053   \\
   165 &   383.7    &   43.6  &  28.4   &  36.0  &  1.6   & 1.19     & 0.14    & 96.0   & 2.22  & 0.023   \\
   170 &   344.0    &   38.5  &  25.3   &  33.4  &  1.4   & 0.79     & 0.09    & 96.5   & 2.36  & 0.016   \\
   175 &   309.7    &   34.0  &  22.5   &  31.0  &  1.3   & 0.61     & 0.07    & 95.8   & 3.23  & 0.012   \\
   180 &   279.2    &   30.1  &  20.0   &  28.7  &  1.1   & 0.50     & 0.06    & 93.2   & 6.02  & 0.010   \\
   185 &   252.1    &   26.9  &  17.9   &  26.9  &  1.0   & 0.39     & 0.05    & 84.4   & 15.0  & 0.008   \\
   190 &   228.0    &   24.0  &  16.1   &  25.1  &  0.9   & 0.32     & 0.04    & 78.6   & 20.9  & 0.007   \\
   195 &   207.2    &   21.4  &  14.4   &  23.3  &  0.8   & 0.27     & 0.03    & 75.7   & 23.9  & 0.006   \\
   200 &   189.1    &   19.1  &  13.0   &  21.7  &  0.7   & 0.24     & 0.03    & 74.1   & 25.6  & 0.005   \\ 
\end{tabular}   
\end{ruledtabular}
\end{center}    
\end{sidewaystable*}

We include all significant Higgs boson production modes in the \hww,
\hzz, and \hgg\ searches.  Besides gluon-gluon fusion through virtual
quark loops ({\it ggH}), we include Higgs boson production in
association with a $W$ or $Z$ vector boson ($VH$) or with a
top-antitop quark pair ($ttH$), and vector boson fusion (VBF). For the
\hbb\ searches, we target the {\it WH}, {\it ZH}, VBF, and $ttH$
production modes with specific searches.  In addition to the leading
signal production mode in each final state, we include contributions of
sub-leading signal production mode, which lead to increased signal
acceptance.  The predictions for the {\it WH} and {\it ZH} cross
sections are taken from Ref.~\cite{djouadibaglio}.  This calculation
starts with the NLO calculation of {\sc v2hv}~\cite{v2hv} and includes
NNLO QCD contributions~\cite{vhnnloqcd}, as well as one-loop
electroweak corrections~\cite{vhewcorr}.  A similar calculation of the
{\it WH} cross section is available in Ref.~\cite{grazziniferrera}.
The VBF cross section is computed at NNLO in QCD in
Ref.~\cite{vbfnnlo}.  Electroweak corrections to the VBF production
cross section are computed with the {\sc hawk} program~\cite{hawk},
and are small and negative (2-3\%) in the Higgs boson mass range
considered here.  We include these corrections in the VBF cross
sections used for this result.  The $ttH$ production cross sections we
use are from Ref.~\cite{tth}.

We use the predictions for the branching ratios of the Higgs boson
decay from Refs.~\cite{lhcxs,lhcdifferential}.  In this calculation,
the partial decay widths for all Higgs boson decays except to pairs of
$W$ and $Z$ bosons are computed with \textsc{HDECAY}~\cite{hdecay},
and the $W$ and $Z$ pair decay widths are computed with {\sc
Prophecy4f}~\cite{prophecy4f}.  The relevant decay branching ratios
are listed in Table~\ref{tab:higgsxsec}.  The uncertainties on the
predicted branching ratios from uncertainties in the charm- and
bottom-quark masses, $\alpha_s$, and missing higher-order effects are
presented in Refs.~\cite{dblittlelhc,denner11}.

\begin{table*}
\caption{\label{tab:cdfacc}Luminosities, explored mass ranges, and references
for the various processes and final states ($\ell$ represents $e$ or $\mu$ and 
$\tau_{\rm{had}}$ denotes a hadronic tau-lepton decay) for combined analyses.  
The generic labels ``$1\times$'', ``$2\times$'', ``$3\times$'', and ``$4\times$'' 
refer to separations based on lepton or photon categories.}
\begin{ruledtabular}
\begin{tabular}{lccc} \\
Channel & Luminosity  & $m_H$ range & Reference \\
        & (fb$^{-1}$) & (GeV/$c^2$) &           \\ \hline
$WH\rightarrow \ell\nu b\bar{b}$ 2-jet channels \ \ \ 4$\times$(5 $b$-tag categories)                                   & 9.45 & 90--150 & \cite{cdfwh} \\
$WH\rightarrow \ell\nu b\bar{b}$ 3-jet channels \ \ \ 3$\times$(2 $b$-tag categories)                                   & 9.45 & 90--150 & \cite{cdfwh} \\
$ZH\rightarrow \nu\bar{\nu} b\bar{b}$ \ \ \ (3 $b$-tag categories)                                                      & 9.45 & 90--150 & \cite{cdfmetbb} \\
$ZH\rightarrow \ell^+\ell^- b\bar{b}$ 2-jet channels \ \ \ 2$\times$(4 $b$-tag categories)                              & 9.45 & 90--150 & \cite{cdfllbb} \\
$ZH\rightarrow \ell^+\ell^- b\bar{b}$ 3-jet channels \ \ \ 2$\times$(4 $b$-tag categories)                              & 9.45 & 90--150 & \cite{cdfllbb} \\
$H\rightarrow W^+ W^-$ \ \ \ 2$\times$(0 jets)+2$\times$(1 jet)+1$\times$(2 or more jets)+1$\times$(low-$m_{\ell\ell}$) & 9.7  & 110--200 & \cite{cdfhww} \\
$H\rightarrow W^+ W^-$ \ \ \ ($e$-$\tau_{\rm{had}}$)+($\mu$-$\tau_{\rm{had}}$)                                          & 9.7  & 130--200 & \cite{cdfhww} \\
$WH\rightarrow WW^+ W^-$ \ \ \ (same-sign leptons)+(tri-leptons)                                                        & 9.7  & 110--200 & \cite{cdfhww} \\
$WH\rightarrow WW^+ W^-$ \ \ \ (tri-leptons with 1 $\tau_{\rm{had}}$)                                                   & 9.7  & 110--200 & \cite{cdfhww} \\
$ZH\rightarrow ZW^+ W^-$ \ \ \ (tri-leptons with 1 jet)+(tri-leptons with 2 or more jets)                               & 9.7  & 110--200 & \cite{cdfhww} \\
$H\rightarrow ZZ$ \ \ \ (4 leptons)                                                                                     & 9.7  & 120--200 & \cite{cdfzz4l} \\
$H \rightarrow \tau^+ \tau^-$ \ \ \ (1 jet)+(2 or more jets)                                                            & 6.0  & 100--150 & \cite{cdftautau} \\
$WH+ZH\rightarrow jjb{\bar{b}}$ \ \ \  (2 $b$-tag categories)                                                           & 9.45 & 100--150 & \cite{cdfjjbb} \\
$H \rightarrow \gamma \gamma$ \ \ \  1$\times$(0 jet)+1$\times$(1 or more jets)+3$\times$(all jets)                     & 10.0 & 100--150 & \cite{cdfhgamgam} \\
$t\bar{t}H \rightarrow W W b\bar{b} b\bar{b}$ \ \ \  (4 jet, 5 jet, $\ge$ 6 jet)$\times$(5 $b$-tag categories)          & 9.45 & 100--150 & \cite{cdftth} \\
\end{tabular}
\end{ruledtabular}
\end{table*}

\section{Search channels}
\label{sec:channels}

Individual searches typically consist of an event selection based on
the topology and kinematic properties of the final state for the
specific Higgs boson production and decay mode under consideration.
Separation of a potential signal from the remaining background
contributions is obtained in most cases by performing a fit, using the
signal and background models, for a single discriminant variable that
is the output of a multivariate algorithm, which considers many
kinematic event variables as its inputs.  The quality of the
background model prediction for the distribution of each input
variable and the final discriminant is studied using orthogonal data
samples carefully selected to validate the modeling of the major
background components of each analysis channel.  The search samples of
each analysis are divided into various sub-channels based on event
information such as types of reconstructed leptons, jet multiplicity,
and $b$-quark-tagging characterization criteria.  A summary of the
individual searches and the sub-channels included within each is given
in Table~\ref{tab:cdfacc}.  We attempt to group events with similar
signal-to-background ratios within individual sub-channels to optimize
search sensitivities.  This approach allows the inclusion of
information from less sensitive event topologies without degrading the
overall sensitivity (for example, events containing higher-impurity
lepton types).  In addition, the isolation of specific signal and
background components within individual sub-channels allows for
further optimization of the multivariate discriminants trained for
each, leading to additional gains in search sensitivity.  The final
multivariate discriminants are separately optimized for each Higgs
boson mass hypothesis in 5~GeV/$c^2$ steps within the mass range under
consideration.

\subsection{$H\rightarrow b{\bar{b}}$ searches}

For searches focusing on the \hbb\ decay, the efficiency for identifying
reconstructed jets originating from $b$ quarks and the resolution of
the invariant-mass measurement from the two $b$-quark jets are of
primary importance.  The three most sensitive searches in this decay
mode utilize a recently-developed multivariate $b$-quark-tagging
algorithm (HOBIT)~\cite{cdfHobit} which is based on the kinematic
properties of tracks associated with displaced decay vertices and
other characteristics of reconstructed jets sensitive to the flavor of
the initiating parton.  {\it Tight} and {\it loose} operating points 
are defined for this algorithm.  For example, the loose operating point 
is found to have a $b$-quark tagging efficiency of approximately 70\% 
and an associated misidentification rate for light quarks and gluons of 
approximately 5\%.  Compared to the SECVTX~\cite{secvtx} algorithm, the 
most commonly used $b$-quark tagging algorithm at CDF, the new algorithm 
improves $b$-tag efficiency by roughly 20\%, for operating points with 
equivalent misidentification (mistag) rates.  The secondary channels 
that require $b$-jet identification (the all-hadronic and
$t{\bar{t}}H\rightarrow t{\bar{t}}b{\bar{b}}$ searches) were not
updated to use the HOBIT tagger and instead rely on the the SECVTX and
JETPROB~\cite{jetprob} algorithms.  The decay width of the Higgs boson
is expected to be much smaller than the experimental dijet mass 
resolution, which is typically 15\% of the mean reconstructed mass.
The \hbb\ searches are most sensitive in final states that include two
jets.  However, sometimes initial-state or final-state radiation can 
produce a third jet in the event.  Including three-jet events increases 
signal acceptance and adds sensitivity, motivating the inclusion of 
these events in the \hbb\ searches.  Since a SM Higgs boson signal would 
appear as a broad enhancement in the reconstructed mass distribution of 
candidate $b$-quark-jet pairs, dedicated efforts to improve the mass 
resolution have been performed in each sub-channel~\cite{HBBJERNN}.  
Along with improved $b$-jet identification and jet-energy resolution, 
the primary \hbb\ analyses have all benefited from increased trigger 
acceptance by including events from many different trigger paths.  In 
many cases, the complicated correlations between kinematic variables 
used in the trigger decision are modeled with a neural network using 
linear regression based on event kinematic properties and geometric 
acceptance~\cite{ZHllbb79}.

\subsubsection{$WH\rightarrow\ell\nu b{\bar{b}}$ search}

The search focusing on the \WH\ production and decay mode~\cite{cdfwh}
has separate analysis channels for events with two and three
reconstructed jets.  Events are further separated into sub-channels
based on the type of reconstructed lepton and the quality of the
tagging information associated with the candidate $b$-quark jets.  
In particular, separate sub-channels are used for events containing 
a high-quality central muon or central electron candidate, a forward
muon candidate, a forward electron candidate, and a looser central 
electron or muon candidate based on the presence of an isolated
track~\cite{WHPRD56,WHPRD75}.  The final two-lepton categories, which
provide some acceptance for lower-quality electrons and single prong
tau decays, are considered only in the case of two-jet events.  For
two-jet events, five sub-channels are used associated with each lepton
category based on the quality of the $b$-tagging information
associated with each jet: two tight tags (TT), one tight and one loose
tag (TL), a single tight tag (Tx), two loose tags (LL), and a single
loose tag (Lx).  In the case of three-jet events, only two $b$-tag
sub-channels, TT and TL, are considered since the other categories
contribute negligibly to the overall search sensitivity.  A Bayesian
neural network is used to distinguish potential Higgs boson signal
events from other background contributions.

\subsubsection{$ZH\rightarrow\ell^+\ell^-b{\bar{b}}$ search}

The search for $ZH\rightarrow\ell^+\ell^-b{\bar{b}}$ production and
decay~\cite{cdfllbb} is based on events with two isolated leptons and
a minimum of two jets.  A combination of triggers based on
electromagnetic energy clusters and signals in the muon chambers matched to
reconstructed tracks are used to to select events containing $Z
\rightarrow ee$ and $Z \rightarrow \mu \mu$ candidates.  Some triggers
based on missing transverse energy requirements are also used to
select $Z \rightarrow \mu \mu$ candidates, taking advantage of the
apparent imbalance in transverse energies that results from the muons
depositing only a small fraction of their energies in the calorimeter.
Neural networks are used to select di-electron and di-muon
candidates~\cite{ZHllbb79}.  The absence of missing energy from
neutrinos allows for improved dijet mass resolution through event-wide
transverse momentum constraints.  These are incorporated through
corrections to the measured jet energies based on the observed \met\
using a neural-network approach.  The search maintains separate
analysis channels for events with two and three jets, as well as for
events with di-electron and di-muon candidates.  Each of the resulting
four channels is further split into four sub-channels (TT, TL, Tx, and
LL) based on the quality of the tagging information available from the
multivariate algorithm.  Signal is separated from background in
multiple phases.  First, three networks are used to distinguish the {\it ZH}
signal from each of the $t\bar{t}b$, $Z+$jets, and diboson
backgrounds.  Then, a final network further separates the signal from all
backgrounds simultaneously.

\subsubsection{\metbb\ search}

The search for \metbb\ production and
decay~\cite{cdfmetbbSecVtx,cdfmetbb} is based on events with large
\met\ and no isolated-lepton candidates.  Additional background
suppression techniques are applied to reduce large background
contributions from multi-jet production processes.  In particular,
prior to construction of the final discriminant, a requirement on a
multivariate discriminant trained specifically for separating
potential signal from the multi-jet background is applied to the event
sample.  Events that do not satisfy this requirement are used to
normalize the remaining multi-jet background contribution, which is
modeled using a mistag rate function for gluon and light-quark jets
measured in data and applied to the untagged jets in data events that
otherwise satisfy the kinematic selection criteria.  One of the inputs
to this multivariate discriminant is a track-based
missing-transverse-momentum calculation that discriminates against
false
\met~\cite{Park:2008}.  A second, final multivariate discriminant
is used to separate the potential Higgs boson signal from the
remaining backgrounds, such as $W$ + heavy flavor jets (where heavy
flavor refers to jets originating from $b$ or $c$ quarks) and
$t{\bar{t}}$ production.  Events with two and three jets are treated
as a single search channel that is subdivided into three sub-channels
(TT, TL, and Tx) based on the quality of tagging information from the
multivariate algorithm.

\subsubsection{All-hadronic search}

The all-hadronic search~\cite{cdfjjbb} focuses on {\it WH},{\it ZH}
and VBF production contributing to the $jjb{\bar{b}}$ final state.  We
use events containing four or five reconstructed jets, at least two of
which have been tagged as $b$-quark candidates based on information
from the previously developed SECVTX~\cite{secvtx} and
JETPROB~\cite{jetprob} algorithms.  The use of these two algorithms
results in two search sub-channels containing events with either two
SECVTX tagged jets (SS) or one SECVTX tagged jet and one JETPROB
tagged jet (SJ).  Large multi-jet background contributions are modeled
from the data by applying a measured mistag probability to the
non-$b$-tagged jets within data events that contain a single
$b$-tagged jet but otherwise satisfy event selection requirements.
The multivariate discriminants used to separate potential signal from
the large background contributions are based on kinematic variable
inputs including some variables developed to distinguish the
reconstructed jets originating of $b$ quarks from those of light
quarks and gluons.

\subsubsection{$t{\bar{t}}H\rightarrow t{\bar{t}}b{\bar{b}}$ search}

The search for $t \bar{t} H \rightarrow t \bar{t} b \bar{b}$
production and decay~\cite{cdftth} is based on events with one
reconstructed lepton, large
\met, and four or more reconstructed jets in which at least two jets are identified 
as $b$-quark candidates based on the SECVTX~\cite{secvtx} or
JETPROB~\cite{jetprob} algorithms.  Events containing four, five, and
six or more jets are analyzed as separate channels, and the events
within each channel are further separated into five sub-channels (SSS,
SSJ, SJJ, SS, and SJ), based on the number of tagged jets and the
algorithms contributing to each tag.  Multivariate discriminant
variables are used to separate potential signal from the dominant
$t\bar{t}$ background contributions.
  
\subsection{$H\rightarrow\tau^+\tau^-$ search}

The search for Higgs bosons decaying to tau lepton
pairs~\cite{cdftautau} incorporates potential contributions from all
four production modes.  The search is based on events containing one
electron or muon candidate and one hadronically-decaying tau-lepton
candidate.  To help reduce $Z/\gamma^{\ast}\rightarrow
\tau^+\tau^-$ background contributions, events are also required to 
contain at least one reconstructed jet.  Events that contain one jet
and two or more jets are treated as independent search sub-channels.
Boosted decision trees are trained for both sub-channels to separate
potential signal events from those associated with each significantly
contributing background production mechanism.  Significant numbers of
background events are removed from the samples by placing lower cuts
on the outputs of each boosted decision tree.  The output of the
boosted decision tree trained to separate potential signal from
$Z/\gamma^{\ast}\rightarrow
\tau^+\tau^-$ background contributions is used as the final discriminating
variable for events surviving all of the selection criteria.

\subsection{$H\rightarrow W^+W^-$ search}

In the search for Higgs bosons decaying to $W$ boson
pairs~\cite{cdfhww} the greatest sensitivity originates from Higgs
bosons produced through gluon fusion; however, the signal
contributions from all four production modes are included.  The
primary search is based on events with two oppositely-charged isolated
leptons and large
\met, focusing on the $H \rightarrow W^+W^-
\rightarrow\ell^+\nu\ell^-\nu$ decay mode.  The presence of neutrinos 
in the final state prevents an accurate reconstruction of the
candidate Higgs boson mass, and separation of a potential signal from
background contributions is based on other kinematic variables.  In
particular, the distribution of angular separations between the
final-state leptons produced in the decays of $W^+W^-$ pairs is
significantly different for pairs originating from a spin-zero
particle, such as the Higgs boson signal, and the major backgrounds.

Events in the primary search are separated into eight sub-channels
based on the types of reconstructed leptons, the number of
reconstructed jets, and the invariant mass of the dilepton pair.  In
the case of events with two electron or muon candidates, separate
analysis channels are used for those events with zero, one, and two or
more reconstructed jets.  This separation helps to isolate potential
signal contributions associated with the four signal production
mechanisms as well as specific background contributions such as
$t\bar{t}$ production, which is dominant for events containing two or
more jets.  Based on this separation, the final multivariate
discriminant used for each channel is optimized, leading to a
significant improvement in the overall search sensitivity.

In the case of events with zero or one reconstructed jet, separate
search sub-channels are used for events containing {\it low-purity}
and {\it high-purity} lepton types.  Events containing forward
electron candidates, for example, have much higher background
contributions from $W$+jets and $W$+$\gamma$ production processes
where a jet or photon mimics the signature of an isolated-lepton
candidate.

A separate search sub-channel is used for events in which the
dilepton mass is smaller than 16~GeV/$c^2$.  The
main background event contribution in this kinematic region originates
from $W$+$\gamma$ production, and additional search sensitivity is
obtained from the use of a separately-trained multivariate
discriminant focused on separating the potential signal from this
particular background.  Two additional search sub-channels are used
for events with one electron or muon candidate and a second,
oppositely-charged hadronically-decaying tau-lepton candidate.  These event
samples contain significant background contributions from $W$+jets and
multi-jet production processes, necessitating the use of independent
search channels.  No further separation of events based on the number
of reconstructed jets is performed within these additional
sub-channels.

Higgs boson production in association with a $W$ or $Z$ boson in
conjunction with the decay $H\rightarrow W^+W^-$ leads to additional
potential signal events in other, more exotic final states.  The
signal contributions are expected to be small, but these final states
contain much smaller contributions from SM background processes.  Hence,
the inclusion of these additional sub-channels improves the overall
search sensitivity.  A search for $W^+H\rightarrow
W^+W^+W^-\rightarrow\ell^+\nu\ell^+\nu jj$ production and decay is
included through a sub-channel focused on events containing two
same-sign, isolated lepton candidates and one or more reconstructed
jets.  Two additional sub-channels are used to search for even smaller
potential signal contributions from $W^+H\rightarrow
W^+W^+W^-\rightarrow \ell^+\nu\ell^+\nu\ell^-\nu$ production and
decay.  These sub-channels, one based entirely on electron and muon
candidates and the other requiring exactly one hadronically-decaying
tau-lepton candidate, focus on events with a total of three isolated lepton
candidates.  In all three sub-channels, the final multivariate
discriminants for separating potential signal from other background
contributions incorporate multiple kinematic event variables including
the observed
\met.  The \met\ provides provides good separation against dominant 
background contributions with misidentified lepton candidates because
of the presence of multiple neutrinos within each signal final state.

Finally, we use events with three isolated lepton candidates to search
for $ZH\rightarrow ZW^+W^-\rightarrow\ell^+\ell^-\ell^+\nu jj$
production and decay.  Three-lepton events that are found to contain a
same-flavor, opposite-sign lepton pair with a reconstructed mass
within 10~GeV/$c^2$ of the $Z$ boson mass are classified into one of
two separate sub-channels based on the presence of one reconstructed
jet or two or more reconstructed jets.  Within the second sub-channel,
all final-state particles from the Higgs boson production and decay
are reconstructed (the transverse momentum components of the neutrino
are obtained from the observed \met) and a reconstructed Higgs boson
mass is used as one of the kinematic input variables to the final
multivariate discriminant.

\subsection{$H\rightarrow ZZ$ search}

The search for Higgs bosons decaying to $Z$ boson pairs~\cite{cdfzz4l}
is based on events with four reconstructed lepton candidates
(electrons or muons).  The selected events consist primarily of the
background from non-resonant diboson production of $Z^{*}/Z$-boson
pairs. A four-lepton invariant mass discriminant is used for
separating the potential Higgs boson signal from the non-resonant $ZZ$
background.  The event \met\ is used as an additional discriminating
variable to improve sensitivity to potential four-lepton event
signal contributions from $ZH\rightarrow
ZW^+W^-\rightarrow\ell^+\ell^-\ell^+\nu
\ell^-\nu$ and $ZH\rightarrow \ell^+\ell^-\tau^+\tau^-$ production and 
decay.

\subsection{$H\rightarrow \gamma\gamma$ search}

The search for Higgs bosons decaying to photon pairs~\cite{cdfhgamgam}
incorporates potential signal contributions from all four Higgs
production mechanisms.  Photon candidates are reconstructed in both
the central and forward calorimeters.  Conversion ($\gamma
\rightarrow e^+e^-$) candidates are also reconstructed in the central calorimeter.  Four search channels based on these
candidate types (central-central, central-forward, central-conversion,
and forward-conversion) are formed from the inclusive diphoton event
sample.  In order to better optimize the most sensitive search
category, central-central events are further separated into two
sub-channels consisting of events with zero reconstructed jets (where
the majority of {\it ggH} events are expected) and one or more
reconstructed jets (where the majority of VH and VBF events are
expected).  For these sub-channels, multivariate discriminants using
the reconstructed diphoton mass and other kinematic event variables as
inputs are used to separate the potential signal from the non-resonant
backgrounds.  In the other analysis channels, the diphoton invariant
mass is used as the sole kinematic discriminant.  For each Higgs boson
mass hypothesis, the signal region is defined to be at least $\pm$2
standard deviations of the expected Higgs boson diphoton mass
resolution. The width of signal windows were taken to be 12~GeV/$c^2$,
16~GeV/$c^2$, and 20~GeV/$c^2$ for mass hypotheses of
100--115~GeV/$c^2$, 120--135~GeV/$c^2$, and 140--150~GeV/$c^2$,
respectively.  The sideband regions around each signal search window
are used to normalize background contributions within the signal
region for all sub-channels, and to validate the background modeling
of the multivariate discriminants for central-central events.

\section{Systematic uncertainties}
\label{sec:systematics}

The Higgs boson signal production rate is expected to be small
compared with the copious backgrounds produced in $p{\bar{p}}$
collisions at Tevatron energies.  Systematic uncertainties associated
with background predictions can be significant relative to expected
signal rates in the highest $s/b$ bins of the discriminant
distributions.  Therefore, it is expected that systematic
uncertainties can have a large impact on search sensitivity.  As an
example, in the case of the search for a 125~\gevcc\ Higgs boson the
inclusion of systematic uncertainties weakens the sensitivity of the
combined analysis by roughly 20\%.  We consider uncertainties that
affect the normalizations as well as those that affect the shapes of
the multivariate discriminants used in the searches.  We refer to
these respectively as {\it rate} and {\it shape} uncertainties.  Some
systematic uncertainties are correlated between analyses, between
sub-channels within an analysis, and between signal and background
predictions within a sub-channel.  The nature of the fits that are
performed requires careful evaluation of the common and independent
sources of systematic uncertainty.  The details of the statistical
treatment of the uncertainties are described in
Sec.~\ref{sec:statistics}.

The most important rate uncertainties in the backgrounds to the
$WH\rightarrow Wb{\bar{b}}$ and $ZH\rightarrow Zb{\bar{b}}$ searches
come from the $W+$jets and $Z$+jets backgrounds.  These uncertainties
are separated into heavy-flavor components and mistags.  The mistags
are calibrated using data control samples.  Because these backgrounds
are calibrated {\it in situ} in events with different selection
requirements from the analysis search region, and because the
differences between the predictions and the true rates may not be the
same between the $W$+jets and $Z$+jets samples, the $W$+heavy flavor
and $Z$+heavy flavor uncertainties are not correlated between
analyses, but are correlated between sub-channels of a single
analysis.  This treatment ensures that we do not use the $Z$+jets
searches to cross-calibrate the backgrounds in the $W$+jets searches
and {\it vice versa}.

The uncertainties on the $b$-tag efficiencies for each $b$-jet
selection requirement are evaluated both for true $b$ jets and for
mistagged jets.  These uncertainties are propagated to each $b$-tag
category.  The resulting uncertainties are treated as correlated
between the signal predictions and the background predictions.  The
uncertainties related to the $b$-tag efficiencies are treated as
correlated between analyses that use the same $b$-tag algorithm.
Similarly, the uncertainty on the total integrated luminosity as
measured by the luminosity monitor is considered correlated among all
signal and background MC-based predictions in all analyses.

We ensure that each analysis uses the same cross section assumptions
and theoretical uncertainties on the prediction for the production of
diboson~\cite{mcfm} ($W^+W^-$, $W^\pm Z$, and $ZZ$),
$t{\bar{t}}$~\cite{tt}, and $s$-channel and $t$-channel single-top
quark~\cite{kidonakis_st} events.  The three diboson processes share
common dependencies on factorization and renormalization scales and
PDFs, so we correlate the uncertainties on all three production modes,
and correlate these uncertainties across all channels.

The jet energy scale is calibrated with experimental data using events
in which a photon recoils from a jet, and events in which a
leptonically-decaying $Z$ boson recoils from a jet~\cite{JES}.  The
associated uncertainties are applied to each analysis. They
change the predicted rates of events passing the respective
selections, largely due to jet $E_T$ requirements, but also
distort the predicted shapes of the distributions of the final
discriminant variables.  Hence, the systematic uncertainty from the
jet energy scale can be further constrained {\it in situ}.  We do not,
however, correlate the jet energy scale uncertainty from one analysis
to another, because the analyses handle jet energies differently, and
accept different fractions of quark and gluon jets in
their respective backgrounds.  For example, the neural-network jet
energy correction technique used in the
$ZH\rightarrow\ell^+\ell^-b{\bar{b}}$ channels may have a different
response to the jet energy mismodeling from the response in the other
$H\rightarrow b{\bar{b}}$ channels.  Uncertainties due to
initial-state and final-state gluon radiation are considered
correlated with each other and across channels.

\section{Statistical methods}
\label{sec:statistics}

The results of the searches in each sub-channel are represented as
distributions of data event counts in intervals (bins) of a final
discriminant variable, which is separately optimized for each
sub-channel at each value of the Higgs boson mass $m_H$.  Along with
the observed data are predictions for each relevant source of
background, each source of signal, and the associated uncertainties.
These uncertainties affect the predicted yields of each component of
signal and background, as well as the differential distributions of
the components in each of the histograms.  We also consider
uncertainties that are uncorrelated from one bin to the next of each
component of the predictions, usually coming from the limited size of
Monte Carlo simulated samples.  The signal-to-background ratios in
most of the channels are of the order of a few percent or less.  The
final discriminant histograms classify events into categories with
different signal-to-background ratios.  Events with higher
discriminant output values populate bins with larger
signal-to-background ratios.

This representation of the search results allows for the extraction of
constraints on both the signal production rates in the decay modes
selected and the background rates and shapes.  Indeed, a major
component of the sensitivity of the search stems from the ability of
the data to constrain the rates and shapes of the major background
sources.  The multivariate discriminants sort the events based on
signal purity.  Typically, the low signal-to-background portions of
the histograms have higher statistics and serve to constrain the
background rates.  The shapes of the predictions for each background
provide the basis by which the background prediction is extrapolated
into the signal-rich region, and the shape uncertainties parametrize
the extrapolation uncertainties.

We use the search results to compute upper limits on the signal rate
for SM Higgs boson production, to determine the best-fit value of the
signal strength and couplings, and to compute $p$-values for purposes
of conducting a hypothesis test where the null hypothesis is that a
Higgs boson signal is absent and the test hypothesis is that a SM
Higgs boson is present with mass $m_H$.  We employ both frequentist
and Bayesian techniques.  The upper limits on Higgs boson production
and the cross section fits are based on a Bayesian calculation
assuming a uniform prior probability density of the signal event rate,
truncated to be non negative.  The $p$-values are computed with a
frequentist method, although the handling of the systematic
uncertainties is Bayesian.  The approach is the same as in
Refs.~\cite{cdfstprl,cdfstprd}.  The likelihood function is a product
over all channels of the Poisson probability of observing the data
given the predictions, which depend on the values of the nuisance
parameters that parametrize the systematic uncertainties.  The
likelihood $L$ is shown below multiplied by the prior probability
density $\pi$,
\begin{widetext}
\begin{equation}
L(R,{\vec{s}},{\vec{b}},{\vec{n}},\vec{\nu})\times\pi({\vec{\nu}}) =
 \prod_{i=1}^{N_C}\prod_{j=1}^{N_{\rm{bins}}}(Rs_{ij}*b_{ij})^{n_{ij}}\frac{e^{-(Rs_{ij}+b_{ij})}}{n_{ij}!}
\times
\prod_{k=1}^{n_{\rm{sys}}}e^{-\nu_k^2/2},
\label{eqn:L}
\end{equation}
\end{widetext}
where the first product is over the number of channels ($N_C$), and
the second product is over histogram bins, each containing $n_{ij}$
events.  The observed number of events in bin $j$ of channel $i$ is
$n_{ij}$.  The SM signal prediction in bin $ij$ is $s_{ij}$, summed
over all production and decay modes contributing to channel $i$, and
$b_{ij}$ is the corresponding background prediction in that bin.  The
predictions $s_{ij}$ and $b_{ij}$ are functions of the nuisance
parameters ${\vec{\nu}}$.  A nuisance parameter $\nu_k$ may affect
many signal and background predictions in a correlated way, such as
the uncertainty on the luminosity; it may distort the distributions of
signal and background predictions, as is the case with jet energy
scales, or it may affect only one bin's prediction of one source of
signal or background, as is the case with Monte Carlo statistical
uncertainties.  The prior probability distributions of the nuisance
parameters are assumed to be independent Gaussians, and the units in
which the nuisance parameters are expressed are in standard deviations
(s.d.) with respect to their nominal values.  The prior distributions
for the nuisance parameters are truncated so that no prediction for
the signal or background in any channel is negative.  The factor $R$
is a simultaneous scaling of all signal components.  Thus, each
combination presented here assumes that the relative ratios of the
contributing Higgs boson production and decay modes are as predicted
by the model under test, within their theoretical uncertainties.  We
therefore present separate combinations assuming the SM, and several
choices of models allowing nonstandard couplings.  We also present
separate measurements of $R$ for channels that are sensitive to one
Higgs boson decay mode at a time.

\begin{figure*} \begin{centering}
\includegraphics[width=0.45\textwidth]{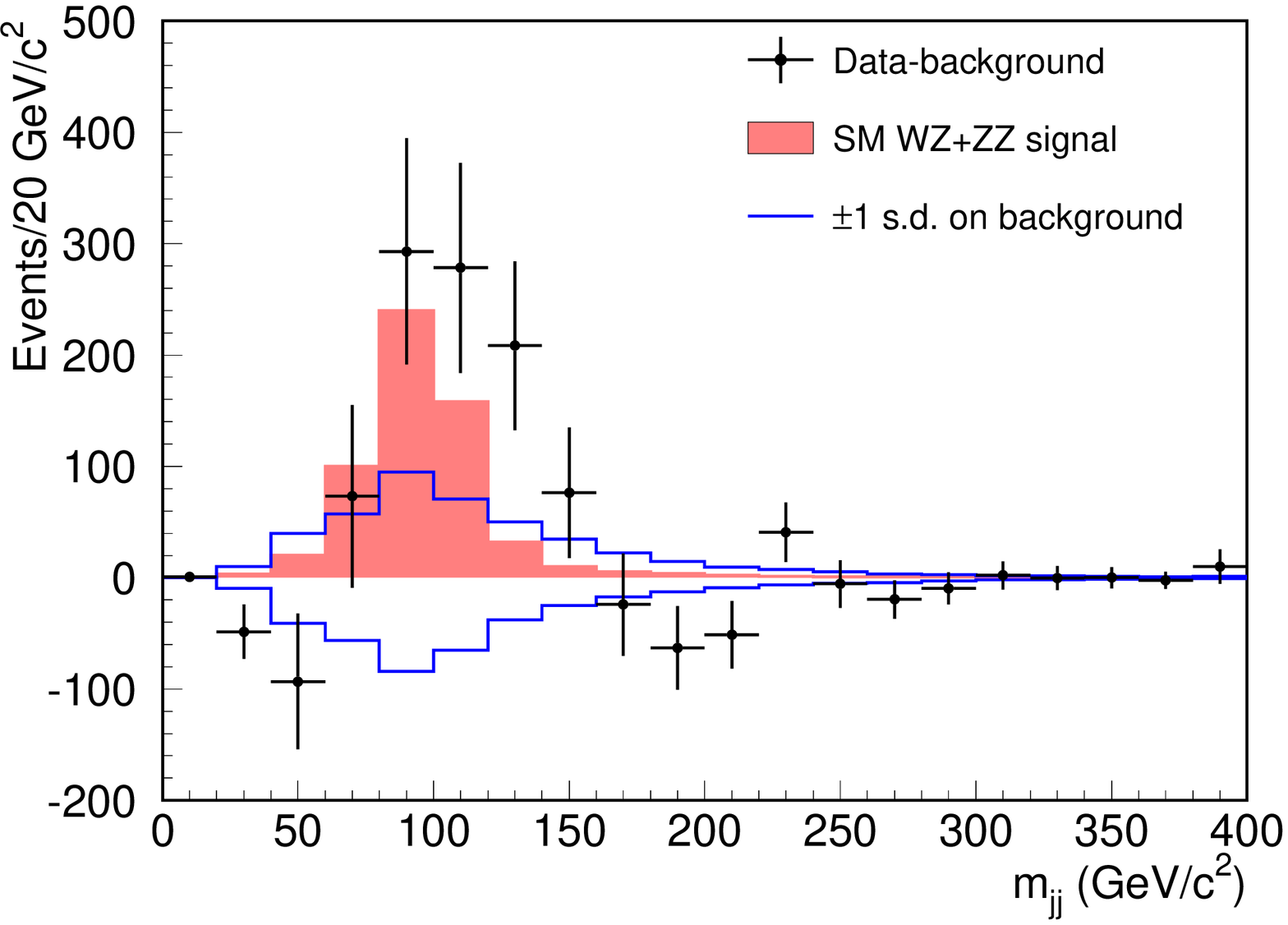}
\includegraphics[width=0.45\textwidth]{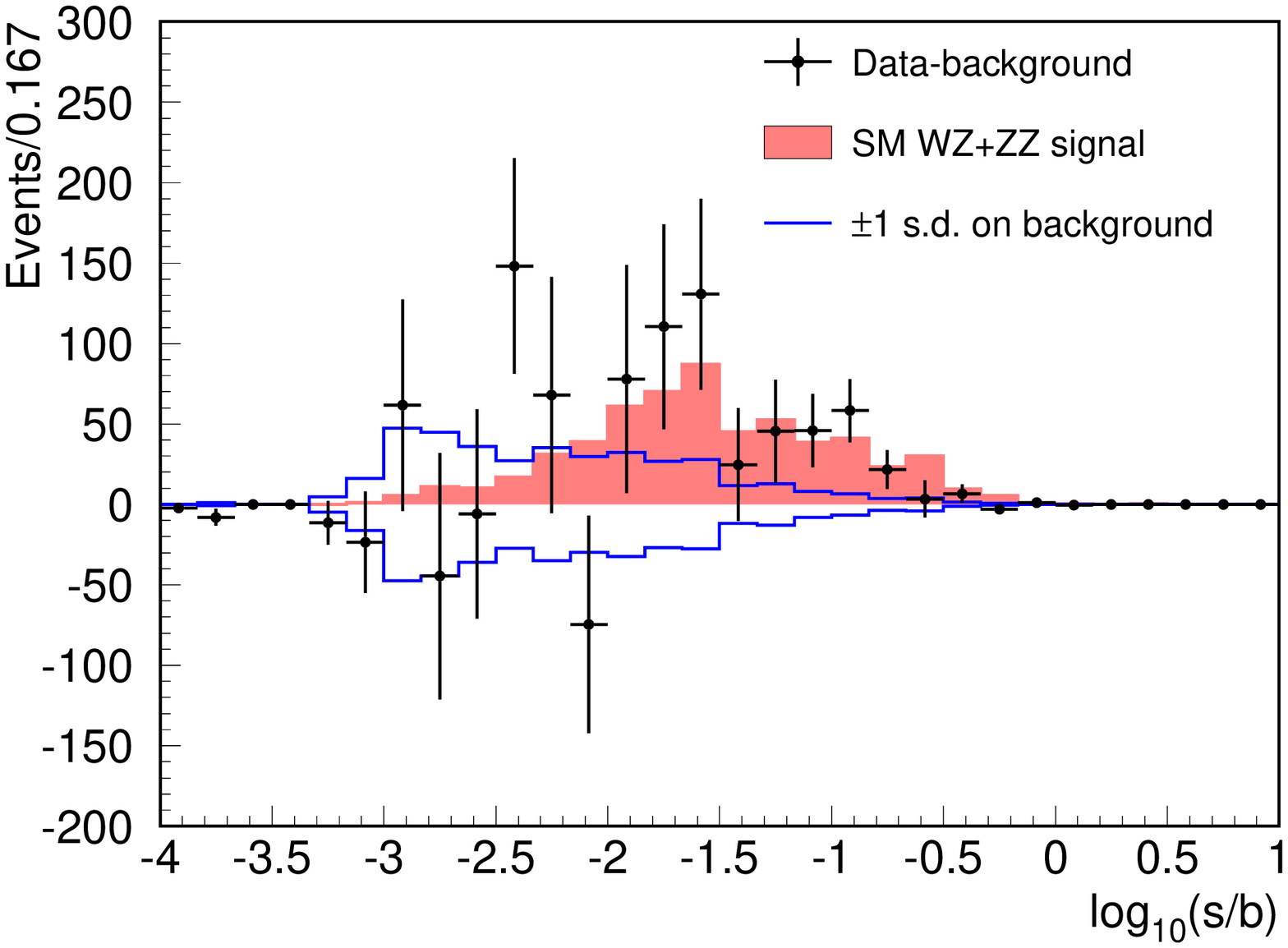}
\caption{
\label{fig:dibo} Background-subtracted dijet invariant mass 
distribution from the combination of all search channels contributing
to the $VZ$ cross section measurement (left) and collected
discriminant histograms, summing bins with similar
signal-to-background ratio ($s/b$), for the $VZ$ measurement (right).
The expected SM signal contributions are indicated with the filled
histograms.  Normalizations of the subtracted background contributions, 
with uncertainties indicated by the unfilled histograms, are obtained 
from fits to the data.}
\end{centering}
\end{figure*}

To calculate the best-fit value of $R$, we assume a uniform prior
probability density $\pi(R)$ for positive values of $R$ and zero for negative
values of $R$, and integrate the likelihood function $L$ multiplied by
the prior probabilities for the nuisance parameters over the values of the nuisance
parameters
\begin{equation}
L^\prime(R,{\vec{s}},{\vec{b}},{\vec{n}}) = \int
L(R,{\vec{s}},{\vec{b}},{\vec{n}},\vec{\nu})\pi({\vec{\nu}})d{\vec{\nu}}.
\end{equation}
The best-fit value of $R$, $R_{\rm{fit}}$, is the value that maximizes
the posterior probability density $L^\prime(R)\pi(R)$.  The 68\% credibility
interval on $R$ is the shortest interval that contains 68\% of the
integral of the posterior density.  We then define the 95\%
credibility upper limit on $R$, $R_{95}$ with the following relation:
\begin{equation}
0.95 = \int_0^{R_{95}} L^\prime(R)\pi(R) dR.
\end{equation}
We compute the distribution of limits that are expected in the
hypothesis that no signal is present by simulating experimental
outcomes and computing $R_{95}$ in each of them.  The experimental
outcomes are simulated by varying the values of the nuisance
parameters within their uncertainties, propagating these to the
predictions of $b_{ij}$, and then drawing simulated data counts from
Poisson distributions with the means of the predicted backgrounds.
The sensitivity is expressed by the median expected limit
$R_{95}^{\rm{med}}$.  A value of $R_{95}<1$ indicates that the
specific signal hypothesis under test is excluded at the 95\%
credibility level.

To evaluate the significance of excess data events compared with the
background prediction, we compute a $p$-value, which is the
probability to observe a result that is as signal-like or more than
the observed result, assuming that no signal is truly present.  A
$p$-value less than $1.35\times 10^{-3}$ is customarily identified as
corresponding to a three s.d. excess, where the correspondence between
the $p$-value and the number of s.d. is computed using the integral of
one tail of a Gaussian distribution.  We rank outcomes as more or less
signal- or background-like based on their $R_{\rm{fit}}$ values.  We
quote the local significance in the SM Higgs boson search at
$m_H=125$~GeV/$c^2$, motivated by the recent discovery by ATLAS and
CMS~\cite{atlasdiscovery,cmsdiscovery}.

\section{Standard model interpretation}
\label{sec:smresults}

\subsection{Diboson production}

The search for the Higgs boson at the Tevatron is challenging due to
large backgrounds relative to the expected signal rate.  Multivariate
techniques are employed to improve sensitivity and this increases the
need to validate the background model predictions for rates and
kinematic distributions.  Over the past few years signals for low
cross section SM processes have been successfully extracted in the
same final states as those used for the primary Higgs boson searches.
For example, the production cross section for electroweak single top
quark production was measured both in the $\ell\nu
b{\bar{b}}$~\cite{cdfstprd} and $\met b\bar{b}$~\cite{stprdmetbb} final
states which provided important validation for the \WH\ and \metbb\
searches.  Similarly, the background model and analysis framework of
the \hww\ search have been validated through successful measurement of
diboson cross sections have been and published in three
final states: $p\bar{p} \rightarrow W^+W^-$ cross section based on the
$\ell^+\bar{\nu}\ell^-\nu$ decay mode~\cite{cdfwwpub}, $p\bar{p}
\rightarrow ZZ$ cross section based on the $\ell^+\ell^-\nu\bar{\nu}$
decay mode~\cite{cdfzzpub}, and a measurement of the $p\bar{p}
\rightarrow W^{\pm}Z$ cross section based on the
$\ell^{\pm}\nu\ell^+\ell^-$ decay mode~\cite{cdfwzpub}.  All three
measurements were found to be in good agreement with NLO predictions.

\begin{figure*} \begin{centering}
\includegraphics[width=0.85\textwidth]{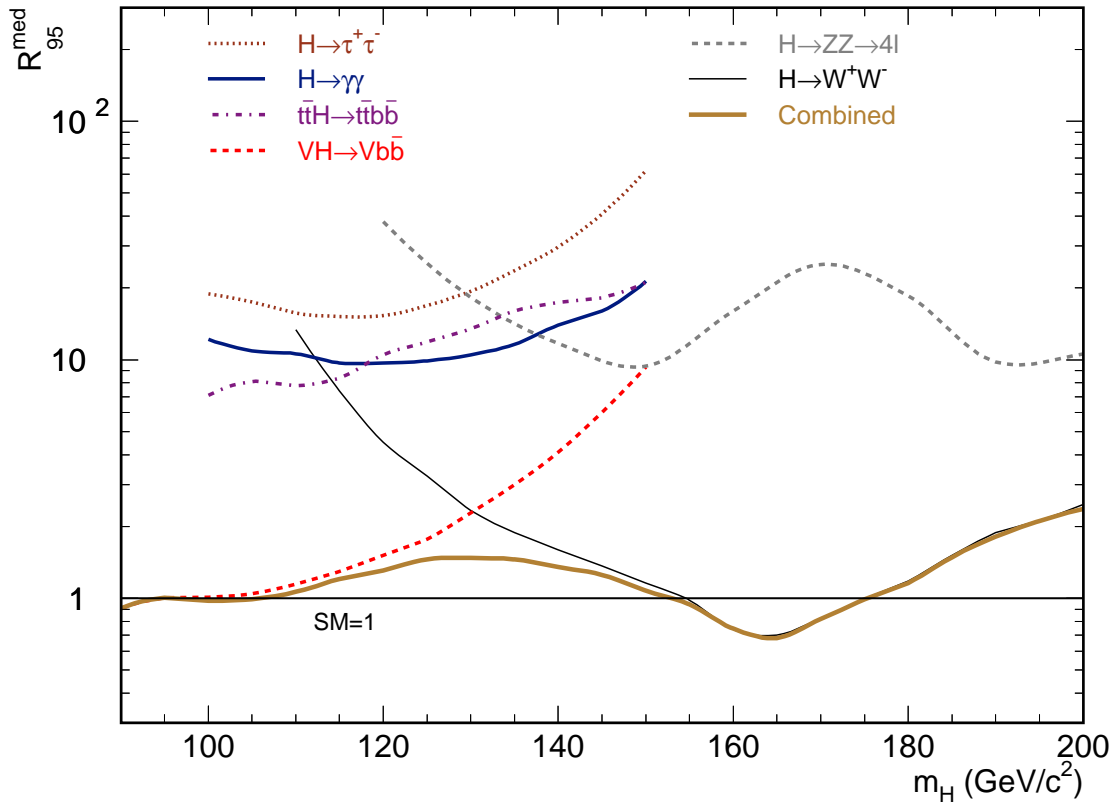}
\caption{
\label{fig:spaghetti} 
Median expected 95\% C.L. upper limits on Higgs boson production
relative to the SM expectation assuming the background-only hypothesis
for combinations of search channels within each Higgs boson decay mode
and the combination of all search channels as a function of Higgs
boson mass in the range between 90 and 200~\gevcc.}
\end{centering}
\end{figure*}

The searches for $WH\rightarrow Wb \bar{b}$ and $ZH\rightarrow Zb \bar{b}$
production and decay require careful modeling of large background
event contributions from $W+$jets and $Z+$jets production.  We gauge 
the sensitivity of these searches, and evaluate the background modeling 
and analysis techniques applied within them, by extracting from these 
search channels a combined cross section measurement for $WZ$ and $ZZ$ 
production.  The NLO SM cross section for $VZ$ production times the 
branching fraction for \Zbb\ is $0.68 \pm 0.05$~pb, about six times 
larger than the expected $0.12 \pm 0.01$ pb cross section times branching 
fraction of $VH \rightarrow Vb\bar{b}$ for a Higgs boson mass of 125~GeV/$c^2$.

This measurement is performed through a combination of the same set of
search channels used for the \WH, \ZH, and \metbb\ Higgs boson
searches.  The data sample, event reconstruction, modeling of signal
and background processes, uncertainties, and sub-channels
are the same as in the Higgs boson search.  However, dedicated
multivariate discriminants are trained to separate event
contributions of $VZ$ production from those of the other backgrounds
and any potential contributions from Higgs boson production are not
considered.  Figure~\ref{fig:dibo} shows the background-subtracted,
reconstructed dijet mass distribution obtained from the
combination of all search channels.  A fit to the data is used to
determine the absolute normalizations for $VZ$ signal and background
contributions.

Separation of the $VZ$ signal component within these search channels
is obtained from a multivariate discriminant that incorporates the
dijet invariant mass as one of its most powerful kinematic inputs.
For improved visualization of the result of the $VZ$ cross section
measurement, we group event counts from all bins of the final
discriminant distributions from each of the search channels with
similar similar signal purity, $s/b$, and display the
background-subtracted data contained within each grouping as a
function of increasing $s/b$ (Fig.~\ref{fig:dibo}).  A fit to the data
is used to determine the absolute normalization of the $VZ$ signal
contribution, indicated by the filled histogram, as well as the
normalization for background contributions. The total uncertainty on
the background prediction is indicated with the unfilled histogram.
Based on the excess of data events in the highest $s/b$ bins, we
measure a $VZ$ production cross section of 2.6 $^{+1.3}_{-1.2}$
(stat.+syst.)~pb, consistent with the SM prediction of $4.4 \pm 0.3$
pb~\cite{mcfm}.

\begin{table*}
\caption{\label{tab:cdfinfo}
Expected number of signal events, Higgs boson mass resolution, and
median expected 95\% C.L. upper limits on Higgs boson production
relative to the SM expectation assuming the background-only hypothesis
for combinations of search channels within each Higgs boson decay mode
at $m_H =$~125~\gevcc. 
$^{\dagger}$Mass resolution is limited in the \hww\ decay mode due to 
the presence of two neutrinos in the final state, which leads to an 
under-constrained system. $^{\ddagger}$Mass resolution is limited in 
the $t\bar{t}H \rightarrow W W b\bar{b} b\bar{b}$ production and decay
mode due to the presence of four $b$ quarks in the final state, which 
leads to an ambiguity in jet assignments for reconstructing the Higgs 
boson mass.}
\begin{ruledtabular}
\begin{tabular}{lccc} \\
Channel & Expected \# of signal events & $m_H$ resolution & Expected
		                              limit \\ & & & relative to SM \\
		                              \hline
\hbb\	                                      & 87.0                         & $\approx$15~\%      & 1.77           \\
\hww\	                                      & 24.2	                     & Limited$^{\dagger}$     & 3.25           \\
$H \rightarrow \gamma \gamma$ & 7.4 & $\approx$2.5~\% & 9.9 \\ 
$t\bar{t}H \rightarrow W W b\bar{b} b\bar{b}$ & 3.6 & Limited$^{\ddagger}$ & 11.9 \\
$H \rightarrow \tau^+ \tau^-$ & 2.3 & $\approx$ 25~\% & 16.9 \\
$H\rightarrow ZZ$ & 0.2 & $\approx$3~\% & 29 \\ 
\end{tabular}
\end{ruledtabular}
\end{table*}

\begin{figure*} \begin{centering}
\includegraphics[width=0.45\textwidth]{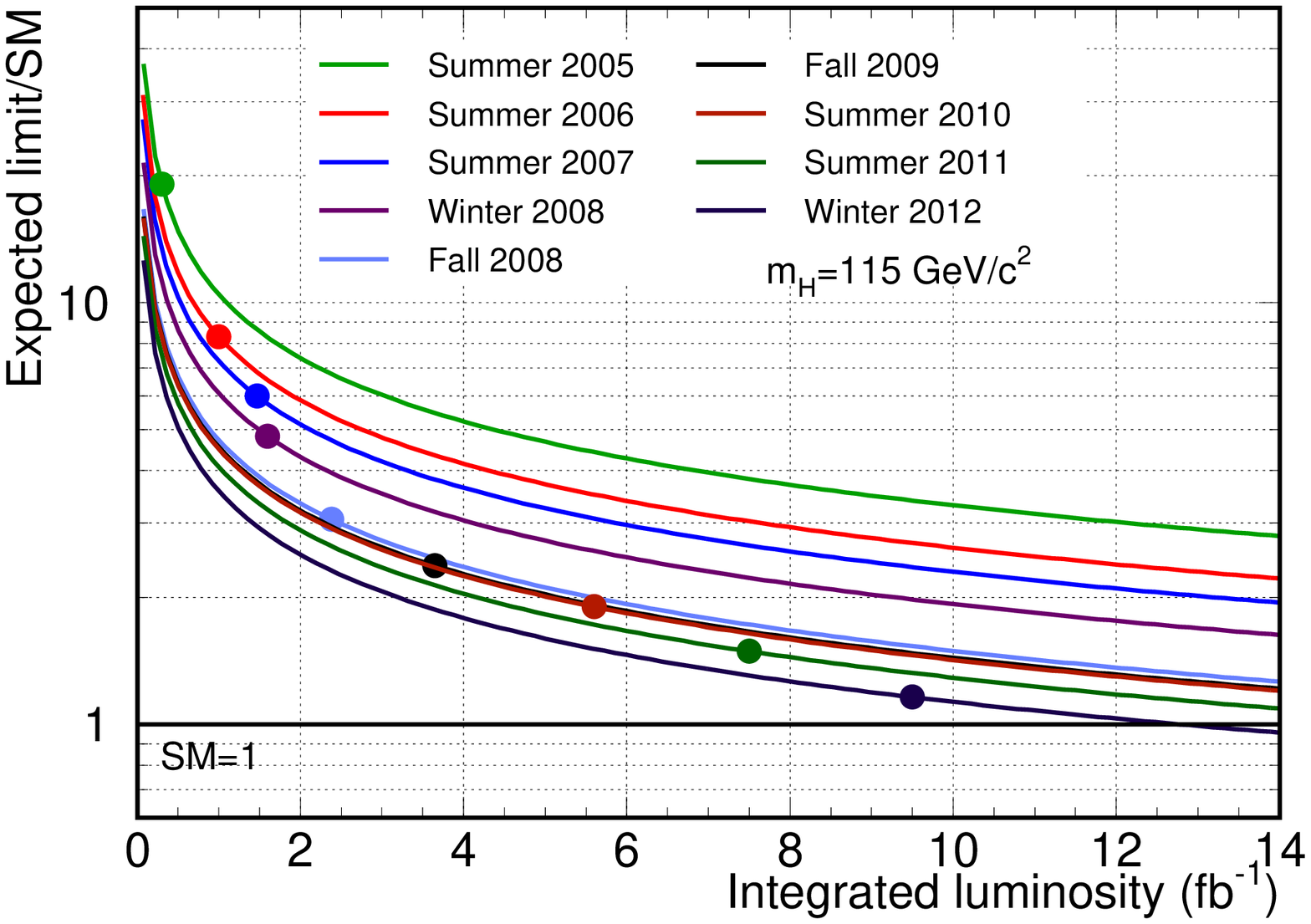}
\includegraphics[width=0.45\textwidth]{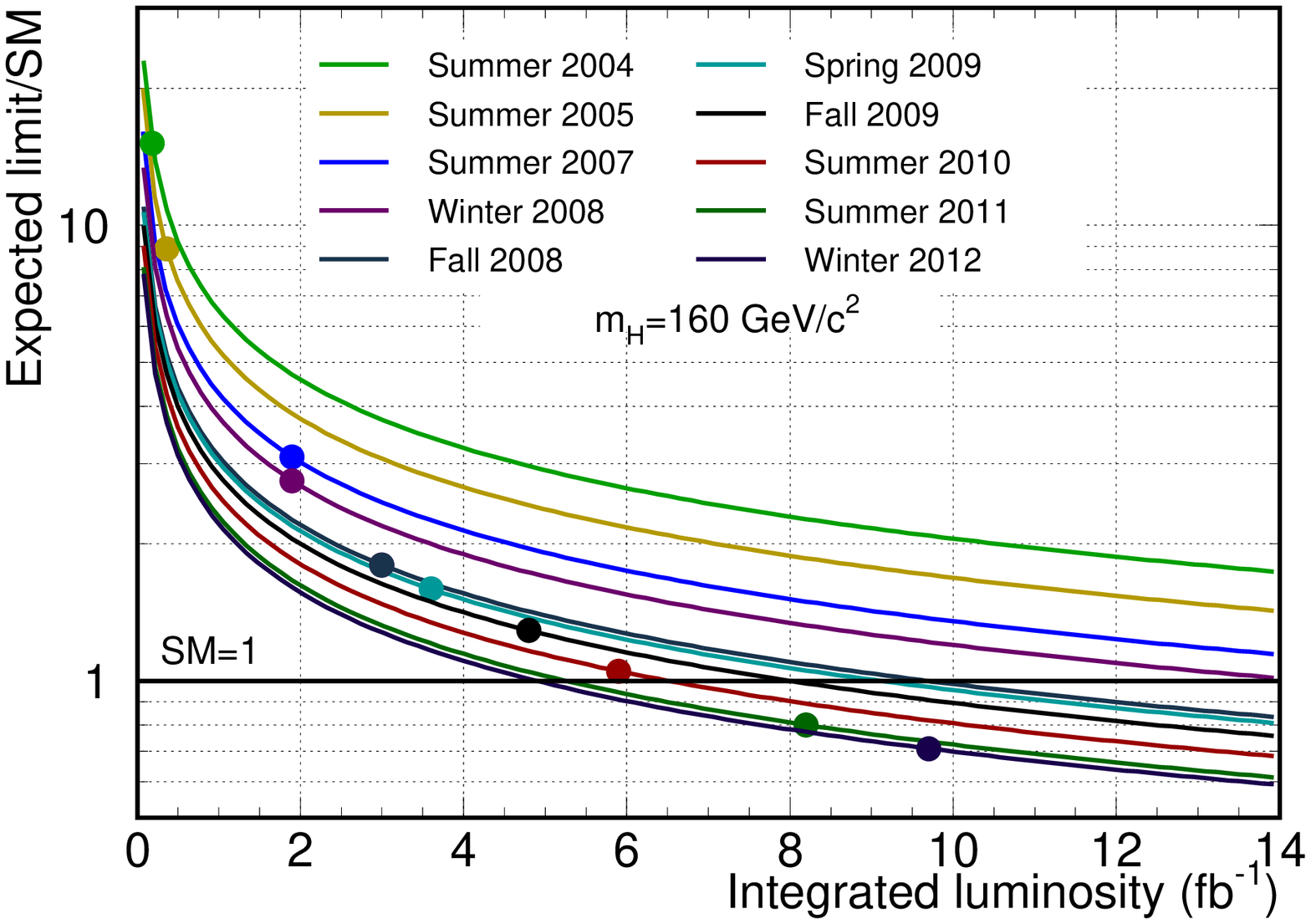}
\caption{
\label{fig:LumiImprove} 
  Achieved and projected median expected 95\% C.L. upper limits on
  Higgs boson production relative to the SM expectation as a function 
  of integrated luminosity, assuming the background-only hypothesis.  
  Each point represents a combination of CDF searches performed on the 
  date indicated in the legend.  The integrated luminosity associated 
  with each point is the sensitivity-weighted average of the analyzed 
  luminosities associated with each contributing channel.  The solid 
  lines show sensitivity projections, where a scaling inversely 
  proportional to the square root of the integrated luminosity is 
  assumed.  The information is provided for combined searches performed 
  at $m_H=115$~\gevcc\ (left) and $m_H=160$~\gevcc\ (right).}
\end{centering}
\end{figure*}

\subsection{Expected sensitivity to Higgs boson production}

The median expected limit in the absence of signal,
$R_{95}^{\rm{med}}$, is shown in Fig.~\ref{fig:spaghetti} for
combinations of the search channels within each Higgs boson decay mode,
and for the full combination of all channels.  For Higgs boson masses
below about 130~\gevcc, searches based on the \hbb\ final state
provide the greatest sensitivity. Searches based on \hww\ are
the most sensitive for higher Higgs boson masses.  Based on the
combined result we expect to exclude the regions
$90<m_H<94$~GeV/$c^2$, $96<m_H<106$~GeV/$c^2$, and
$153<m_H<175$~GeV/$c^2$ in the absence of signal.  For the case of a
Higgs boson with a mass of 125~\gevcc, the signal event yields,
approximate mass resolutions, and median expected limits are shown in
Table~\ref{tab:cdfinfo} for combinations of the channels associated
with each Higgs boson decay mode.  At this mass, \hbb\ has the best
sensitivity, but the \hww\ searches make an important contribution to
the combination.

\begin{figure*} \begin{centering}
\includegraphics[width=0.45\textwidth]{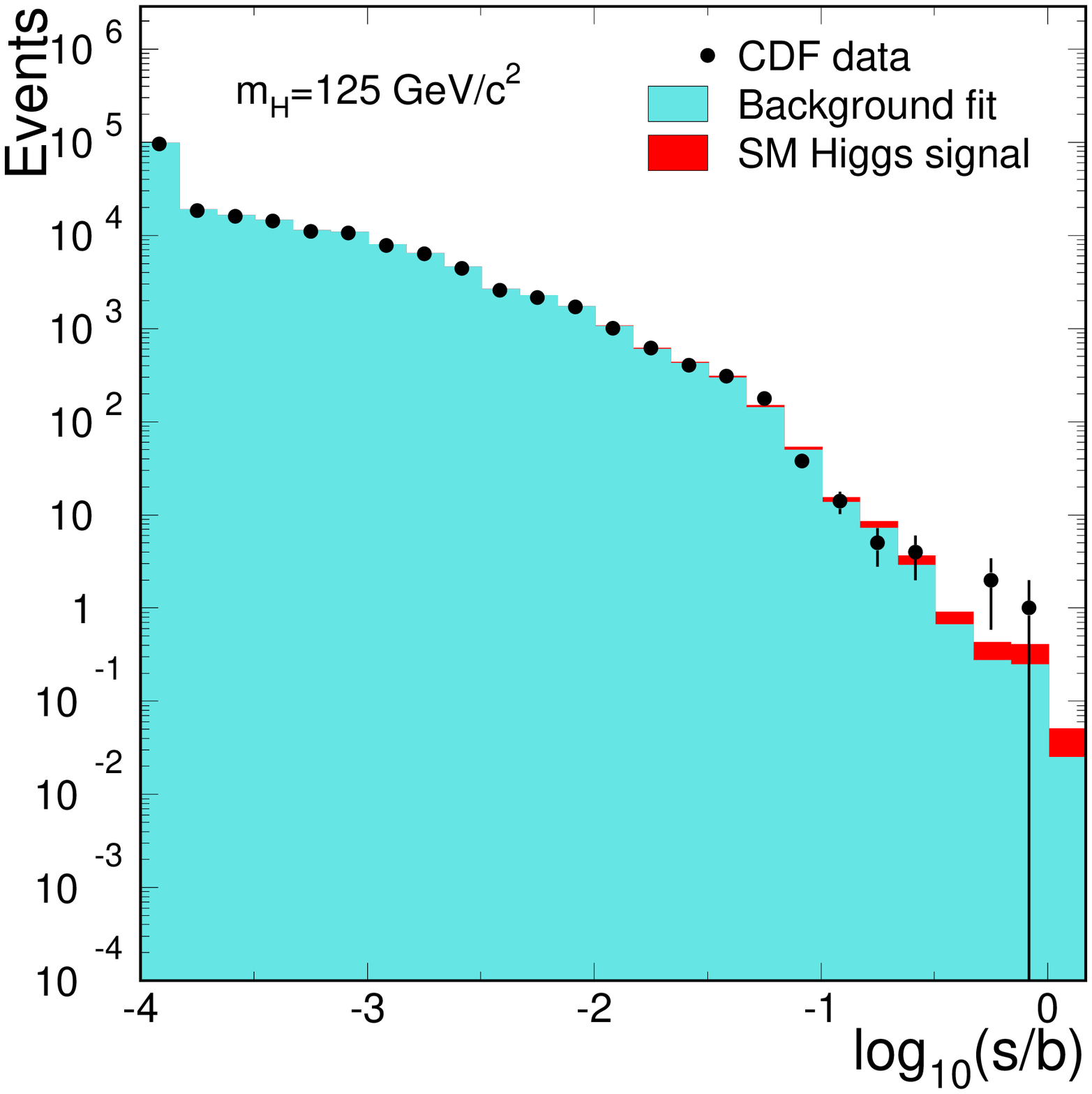}
\includegraphics[width=0.45\textwidth]{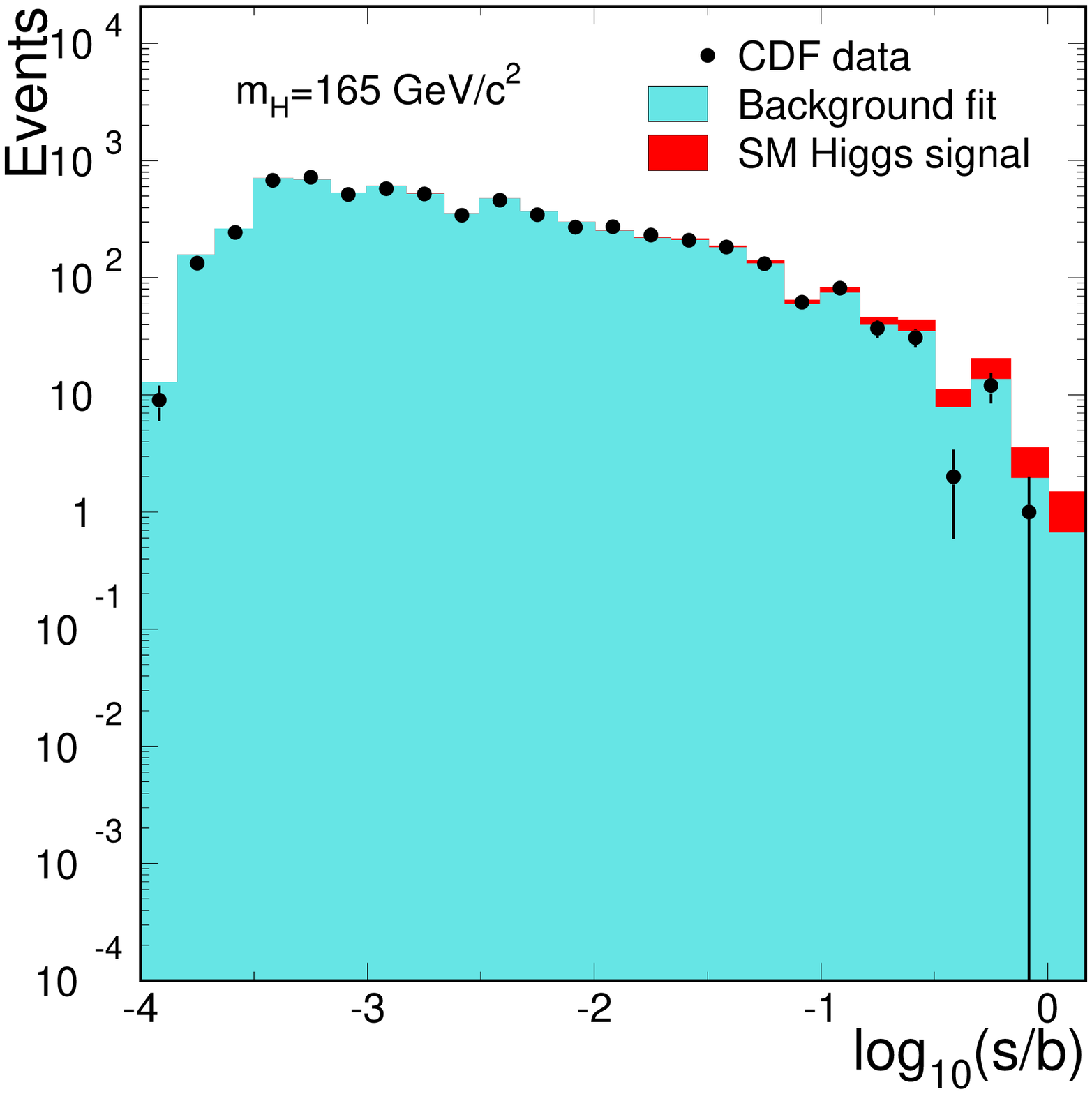}
\caption{
\label{fig:sbdist} Collected discriminant histograms, summed 
for bins with similar signal-to-background ratio ($s/b$), for the
combined SM Higgs boson searches focusing on the $m_H=125$~GeV/$c^2$
(left) and $m_H=165$~GeV/$c^2$ (right) hypotheses.  Normalizations of
the background contributions are obtained from fits to the data.
Predicted signal contributions, scaled to SM expectations, are
overlaid on the backgrounds.}
\end{centering}
\end{figure*}

\begin{figure*} \begin{centering}
\includegraphics[width=0.45\textwidth]{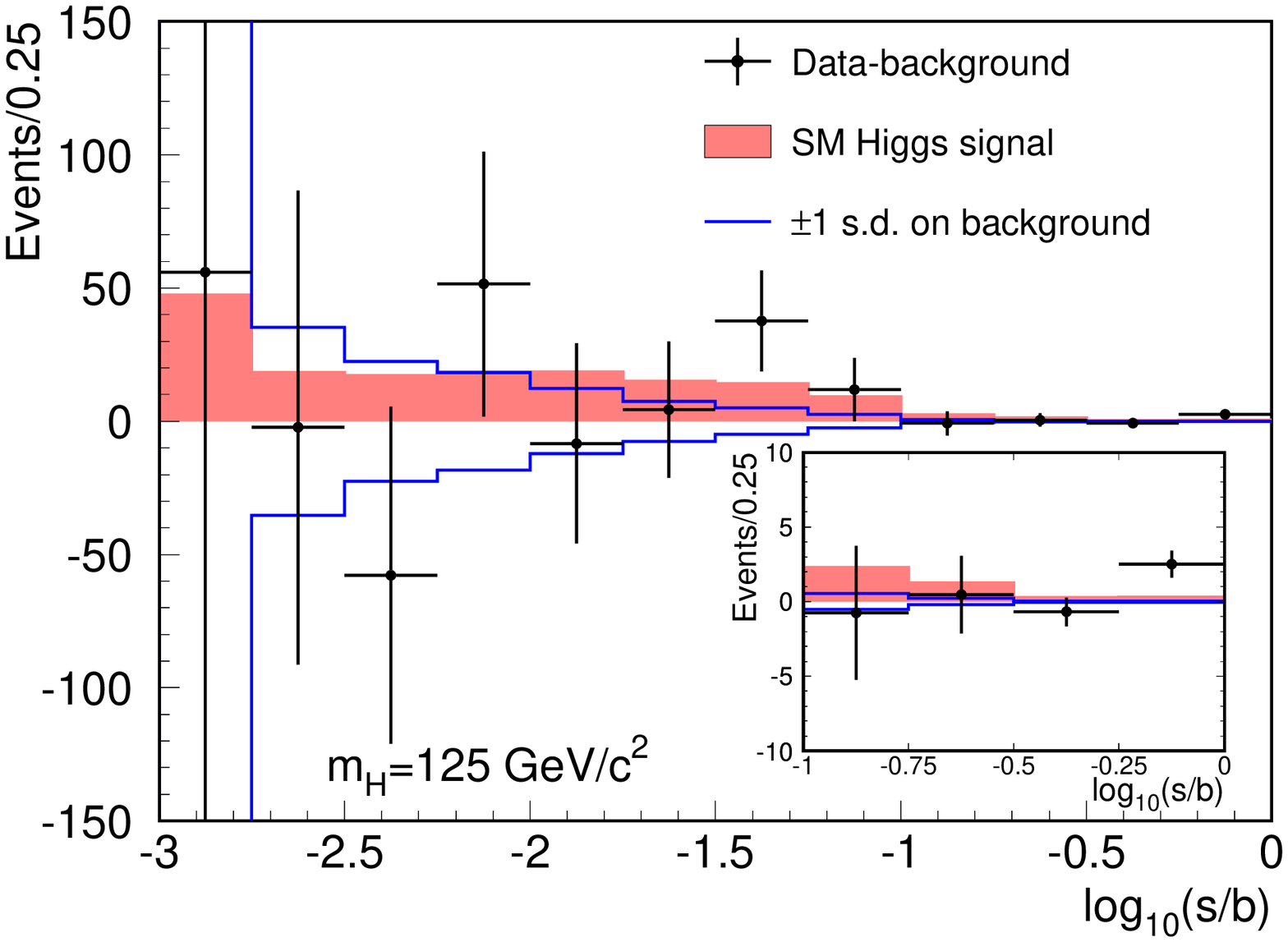}
\includegraphics[width=0.45\textwidth]{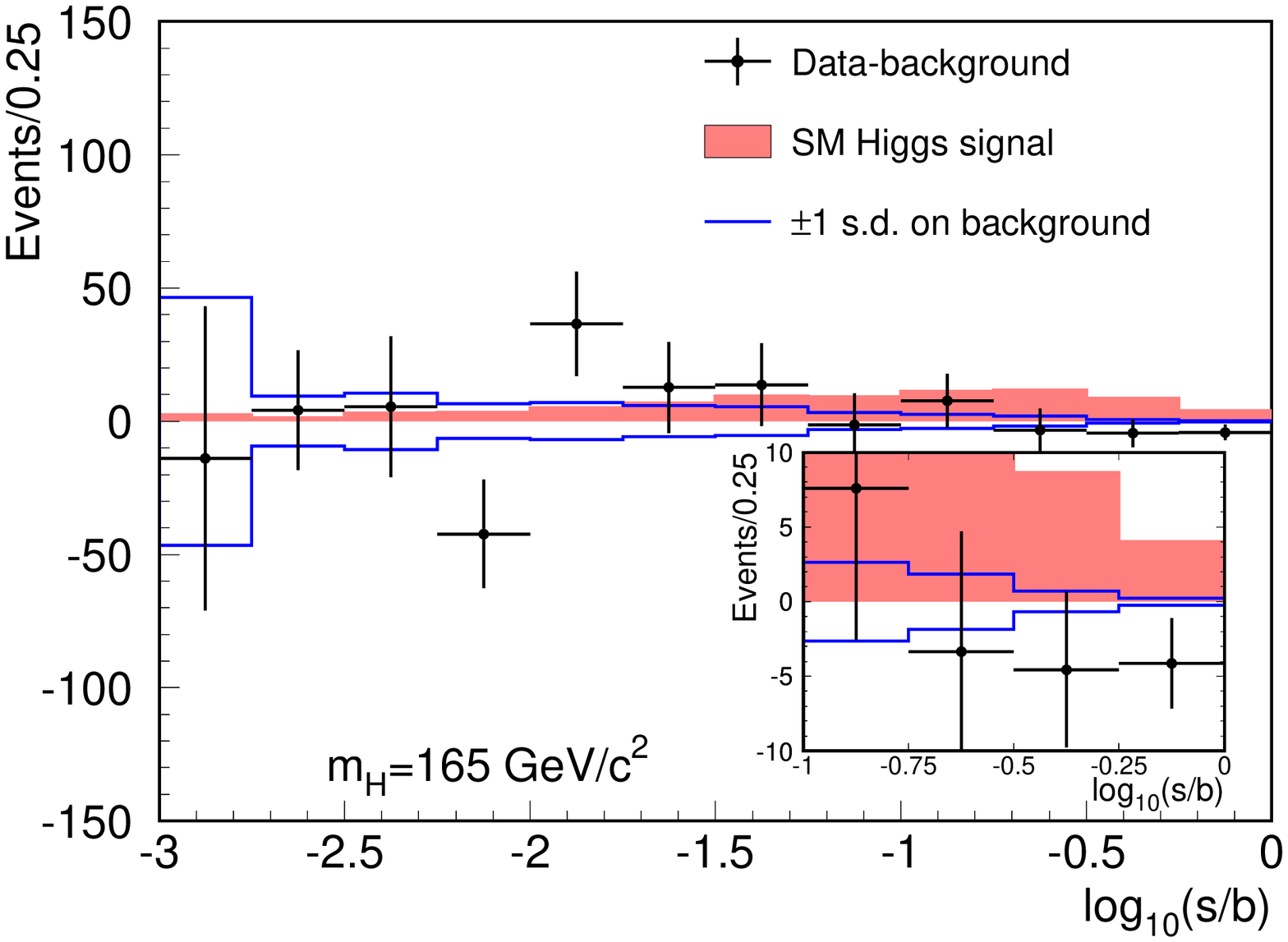}
\caption{
\label{fig:bgsub} Background-subtracted collected discriminant
histograms, summed for bins with similar signal-to-background ratio
($s/b$), for the combined SM Higgs boson searches focusing on the
$m_H=125$~GeV/$c^2$ (left) and $m_H=165$~GeV/$c^2$ (right) hypotheses.
Background normalizations are obtained from fits to the data, and fit
uncertainties are indicated by the unfilled histograms.  Predicted
signal contributions, scaled to SM expectations, are shown with the
filled histograms.  Uncertainties on the data points correspond to the
square root of the sum of expected signal and background yields within
each bin.}
\end{centering}
\end{figure*}

The final sensitivities of CDF Higgs boson searches are a
direct result of a substantial effort made over the last decade to
significantly improve the analysis techniques used.  The
evolution of CDF search sensitivity over time is illustrated in
Fig.~\ref{fig:LumiImprove}.  The points show the median expected 95\%
C.L. upper limits on Higgs boson production relative to SM expectations
assuming the background-only hypothesis from the combination of
available CDF search results performed at various stages over the past
decade.  The integrated luminosities associated with each point are
the sensitivity-weighted averages of analyzed luminosities corresponding 
to the analyzed samples at that time.  The curves show
how the sensitivity of each combination would be expected to improve
in the absence of further analysis improvements assuming that
sensitivity scales inversely with the square root of integrated
luminosity.  With respect to early versions of the CDF Higgs boson
search sensitivity has been improved by more than a factor of two
over what would be expected simply by incorporating more data.  The
illustrated gains in search sensitivity have originated from a
wide array of analysis improvements including the inclusion of
additional triggered events, improved $b$-jet identification
algorithms, implementation of algorithms for improved jet energy
resolution, inclusion of new search channels such as those considering
events with additional jets, and improved multivariate techniques for
separating signal and background contributions.

\subsection{Full combination}

The data are categorized into 81 sub-channels for
the $m_H=125$~GeV/$c^2$ hypothesis.  In order to better visualize
the results and identify data events causing fluctuations in the
observed limits and $p$-values with respect to expectations for the
background-only scenario, we perform a joint fit of the background
predictions for all channels to the observed data where nuisance
parameters are allowed to float within their uncertainties.  We then
collect bins from the final discriminant distribution by merging bins 
with similar $s/b$.  The result is shown in Fig.~\ref{fig:sbdist}, 
for the combined channels contributing to the searches focusing on 
the $m_H=125$~GeV/$c^2$ and 165~GeV/$c^2$ mass hypotheses.  The 
predicted Higgs boson contributions based on SM expectations summed 
over the bins with similar $s/b$ are shown with the fitted background 
contributions overlaid.  A subset of the same data is shown in 
Fig.~\ref{fig:bgsub} where the data are grouped into wider $s/b$ 
bins and the backgrounds determined from the fit have been subtracted.  
A mild excess of data events is observed in the bins with the highest 
$s/b$ for the $m_H=125$~GeV/$c^2$ hypothesis. No such excess is seen 
for the case of the $m_H=165$~GeV/$c^2$ hypothesis.

\begin{figure}[htb] \begin{centering}
\includegraphics[width=0.95\columnwidth]{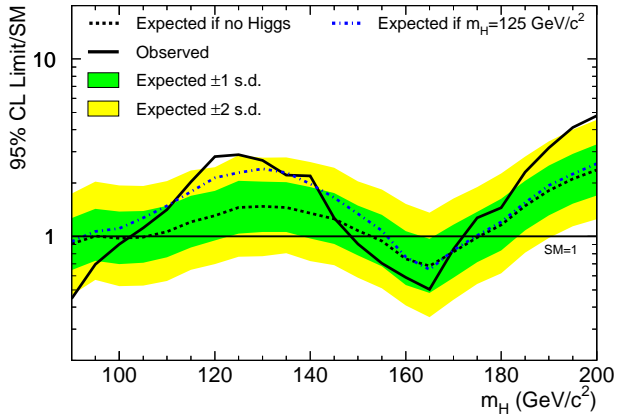}
\caption{
\label{fig:smlimits} 
Observed and expected (median, for the background-only hypothesis)
95\% C.L. upper limits on SM Higgs boson production as a function of
the Higgs boson mass for the combination of CDF searches.  The limits
are expressed as multiples of the SM prediction for test masses in 5
GeV/$c^2$ steps from 90 to 200 GeV/$c^2$.  The points are connected
with straight lines for improved readability. The bands indicate the
68\% and 95\% probability regions where the limits can fluctuate, in
the absence of signal.  The lighter dashed line indicates mean
expected limits in the presence of a SM Higgs boson with $m_H = 125$
GeV/$c^2$.}
\end{centering}
\end{figure}

The likelihood from Equation~\ref{eqn:L} is used to combine the Higgs
boson searches from all CDF sub-channels as described in
Sec.~\ref{sec:statistics}.  Figure~\ref{fig:smlimits} shows the
resulting observed upper bound on the signal scale factor $R_{95}$ for
potential $m_H$ values between 90~GeV/$c^2$ and 200~GeV/$c^2$.  The
median expected limit in the presence of no signal,
$R_{95}^{\rm{med}}$, is shown by the dark dashed line, while the
shaded regions indicate the limit fluctuation ranges at the level of one and
two standard deviations.  The lighter dashed line shows the broad
excess in the limits that would be expected if a SM Higgs boson with
$m_H=125$~GeV/$c^2$ were present in the data.  Values of the observed
and expected limits are listed in Table~\ref{tab:smlimits}.  We
exclude at the 95\% credibility level (C.L.) the SM Higgs boson within
the mass ranges $90<m_H<102$~GeV/$c^2$ and $149<m_H<172$~GeV/$c^2$.
In the absence of a signal, we expect to exclude the regions
$90<m_H<94$~GeV/$c^2$, $96<m_H<106$~GeV/$c^2$, and
$153<m_H<175$~GeV/$c^2$.

\begin{table*}
\caption{\label{tab:smlimits} Median expected (for the background-only 
hypothesis) and observed 95\% C.L. upper limits on Higgs boson
production relative to SM expectations as a function of Higgs boson
mass in GeV/$c^2$ for the combination of CDF searches.}
\begin{ruledtabular}
\begin{tabular}{lcccccccccccc}
Mass & 90 & 95 & 100 & 105 & 110 & 115 & 120 & 125 & 130 & 135 & 140 &
145 \\ \hline
Expected & 0.91 & 1.01 & 0.98 & 0.99 & 1.06 & 1.21 & 1.31 & 1.46 &
1.48 & 1.45 & 1.35 & 1.25 \\ Observed & 0.45 & 0.70 & 0.90 & 1.12 &
1.42 & 2.03 & 2.82 & 2.89 & 2.68 & 2.22 & 2.19 & 1.27 \\
\hline\hline 
Mass & 150 & 155 & 160 & 165 & 170 & 175 & 180 & 185 & 190 & 195 & 200
\\ \hline
Expected & 1.08 & 0.94 & 0.75 & 0.68 & 0.82 & 0.99 & 1.16 & 1.49 &
1.82 & 2.11 & 2.37 \\ Observed & 0.91 & 0.71 & 0.59 & 0.50 & 0.85 &
1.28 & 1.45 & 2.31 & 3.16 & 4.12 & 4.79 \\
\end{tabular}
\end{ruledtabular}
\end{table*}

\begin{figure}[htb] \begin{centering}
\includegraphics[width=0.95\columnwidth]{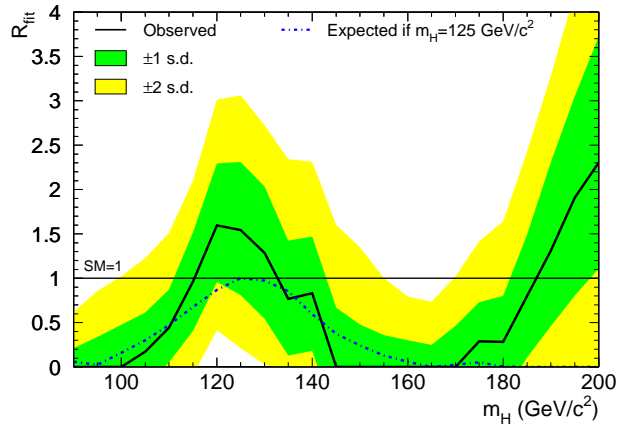}
\caption{\label{fig:smxs} 
Best-fit cross section for inclusive Higgs boson production,
normalized to the SM expectation, for the combination of all CDF
search channels as a function of the Higgs boson mass.  The solid line indicates the fitted cross section,
and the associated shaded regions show the 68\% and 95\% credibility
intervals, which include both statistical and systematic
uncertainties.  The mean expected cross section fit values assuming
the presence of a SM Higgs boson at $m_H=125$~\gevcc\ are shown with
the dot-dashed line.}
\end{centering}
\end{figure}

\begin{figure}[htb] \begin{centering}
\includegraphics[width=0.95\columnwidth]{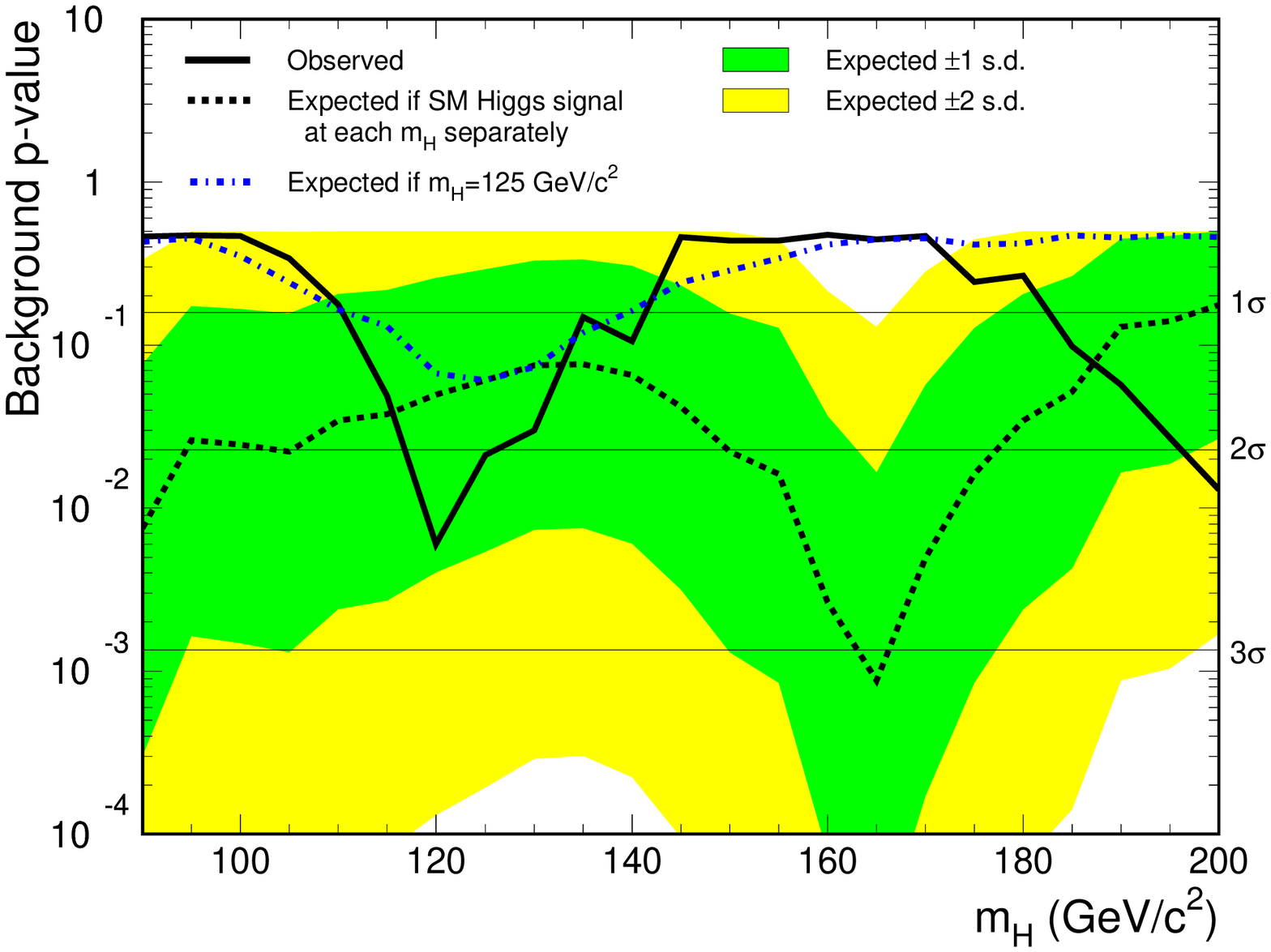}
\caption{\label{fig:smpvalue}  
The significance of the observed data excess with respect to the
background-only expectation for the combination of all CDF search
channels as a function of SM Higgs boson mass.  The probabilities for
the background model to result in a best-fit cross section as large or
larger than that observed in data, $p$-values, are shown with the
solid line.  The dashed line indicates the mean expected $p$-values in
the presence of a SM Higgs boson evaluated separately for each test
mass, where the associated shaded regions show the ranges of one and
two standard deviation fluctuations in observed $p$-values for these
scenarios.  The dot-dashed line indicates mean expected $p$-values for
each mass hypothesis in the case of the SM Higgs boson with
$m_H=125$~GeV/$c^2$.}
\end{centering}
\end{figure}

Mild excesses in the data compared with fitted background predictions are
observed, in particular within the high $s/b$ bins of the
discriminants associated with the $WH\rightarrow\ell\nu b{\bar{b}}$
and $ZH\rightarrow \ell^+\ell^- b{\bar{b}}$ searches~\cite{cdfwh,
cdfllbb}.  However, in the low mass search region where there is
overlap with the $H\rightarrow b\bar{b}$ searches, the $H\rightarrow
W^+W^-$ search, which contributes similar search sensitivity at
$m_H=125$~GeV/$c^2$, does not contain data excesses in the
high $s/b$ bins of its discriminants~\cite{cdfhww}.  By combining
channels, the location of the data excess within the range of
potential $m_H$ values can be partially constrained based on knowledge
of the available mass resolution and expected signal rates from each
search channel.  The constraints are observable in the measured values
for $R_{\rm{fit}}$, which are shown as a function of $m_H$ along with
their associated 68\% and 95\% C.L. intervals in Fig.~\ref{fig:smxs}.  
The  moderate excess is localized within the region 110~$< m_H < 
140$~GeV/$c^2$, where the measured signal rate is found to be 
consistent with that expected from SM Higgs boson production.  The 
best-fit value measured for the Higgs boson production cross section 
at $m_H=125$~GeV/$c^2$ is $1.54^{+0.77}_{-0.73}$ (stat.+syst.) relative 
to the SM prediction.  

The $p$-value is shown as a function of $m_H$ in
Fig.~\ref{fig:smpvalue}.  The broad excess observed in the cross
section measurement is also visible in the $p$-value.  The $p$-value
for the $m_H=125$~GeV/$c^2$ hypothesis is 0.0212 corresponding to a
2.0 standard deviation excess.  A lower $p$-value (0.0060) is observed
for the $m_H=120$~GeV/$c^2$ mass hypothesis, which is not expected to
be distinguishable from the $m_H=125$~GeV/$c^2$ hypothesis based on
the mass resolution of the most sensitive search channels.  There is 
also approximately a two sigma excess in our data for Higgs boson 
mass hypotheses above $\approx$195~\gevcc.  Recent results from the 
LHC experiments strongly exclude the SM Higgs boson in this mass 
range~\cite{atlascombo,cmscombo}.  Taking this into consideration, 
the mild excess near 200~\gevcc\ is likely the result of a statistical 
fluctuation.

\begin{figure}[htb] \begin{centering}
\includegraphics[width=0.95\columnwidth]{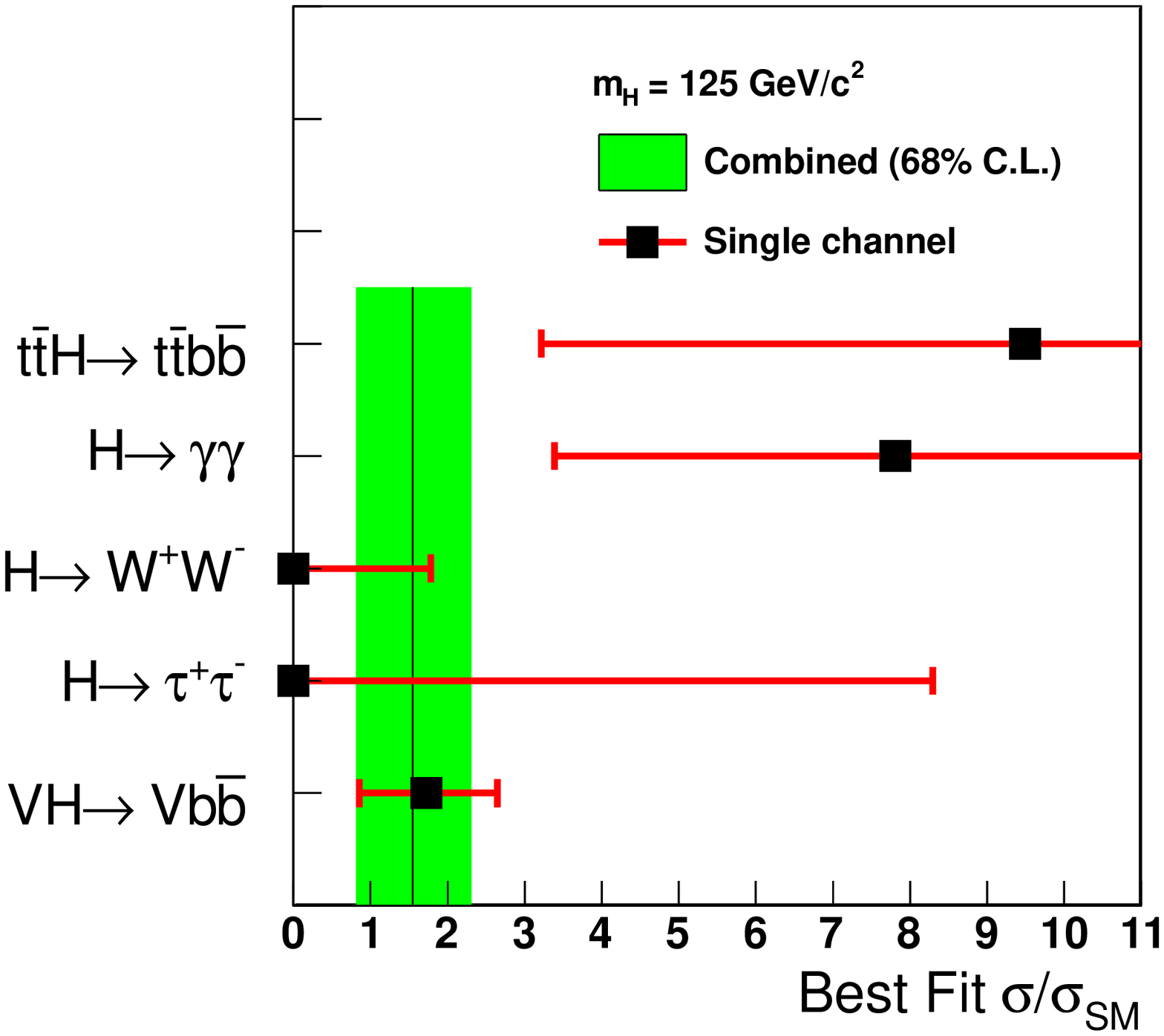}
\caption{
\label{fig:xsecsummary125} Summary of best-fit signal cross sections 
relative to SM expectations for the $m_H=125$~GeV/$c^2$ hypothesis.
Square dots with horizontal uncertainty bars show the fitted cross
sections obtained from the subsets of CDF search channels
corresponding to $VH\rightarrow Vb\bar{b}$, $H\rightarrow W^+W^-$,
$H\rightarrow\gamma
\gamma$, $H\rightarrow \tau^+\tau^-$, and $t\bar{t}H\rightarrow t\bar{t}
b\bar{b}$ production and decay.  The solid vertical line and
associated shaded region illustrate the fitted SM cross section
obtained from all search channels.}
\end{centering}
\end{figure}

We study the couplings of a potential SM Higgs boson by also
extracting best-fit signal cross sections for different combinations
of channels corresponding to specific Higgs boson
production and decay modes.  In particular, we perform cross section
fits for the subsets of CDF search channels corresponding to
$VH\rightarrow Vb\bar{b}$, $H\rightarrow W^+W^-$,
$H\rightarrow\gamma\gamma$, $H\rightarrow \tau^+\tau^-$, and
$t\bar{t}H\rightarrow t\bar{t}b\bar{b}$ production and decay.
Best-fit cross sections relative to SM expectations are provided as a
function of $m_H$ for each of these modes in Table~\ref{tab:xsmeas}.
A comparison of the individual mode fitted cross sections versus the
fitted SM cross section obtained from all search channels is shown in
Fig.~\ref{fig:xsecsummary125} for the $m_H=125$~GeV/$c^2$ hypothesis.
The fitted signal contribution from the $H\rightarrow W^+W^-$ and
$H\rightarrow\tau^+\tau^-$ channels is zero and for the $VH\rightarrow 
b\bar{b}$, $H\rightarrow\gamma\gamma$, and $t\bar{t}H\rightarrow 
t\bar{t}b\bar{b}$ channels it exceeds the SM expectation.  However, 
all best-fit cross sections are found to be consistent within 1.5 
standard deviations of SM Higgs boson expectations.  

\begin{table*}
\begin{center}
\caption{\label{tab:xsmeas} Best-fit signal cross sections, $R_{\rm{fit}}$, as a function 
of $m_H$ for the combination of all SM search channels and for combinations of subsets of 
search channels corresponding to $VH\rightarrow Vb\bar{b}$, $H\rightarrow W^+W^-$, $H
\rightarrow\gamma\gamma$, $H\rightarrow \tau^+\tau^-$, and $t\bar{t}H\rightarrow t\bar{t}
b\bar{b}$ production and decay.  The quoted uncertainties bound the smallest interval containing 
68\% of an integral over the posterior probability densities, which include both statistical and systematic effects.}
\begin{ruledtabular}
 \begin{tabular}{ccccccc}
 $m_H$       & $R_{\rm{fit}}$           & $R_{\rm{fit}}$           & $R_{\rm{fit}}$            & $R_{\rm{fit}}$             & $R_{\rm{fit}}$             & $R_{\rm{fit}}$                                \\
 (GeV/$c^2$) & SM combination           & $H\rightarrow W^+W^-$    & $H\rightarrow b{\bar{b}}$ & $H\rightarrow\gamma\gamma$ & $H\rightarrow\tau^+\tau^-$ & $t{\bar{t}}H\rightarrow t{\bar{t}}b{\bar{b}}$ \\
 \hline
 90          & $0.00 ^{+0.21 }_{-0.00}$ &                          & $0.00 ^{+0.21 }_{-0.00}$  &                            &                            &                                               \\
 95          & $0.00 ^{+0.34 }_{-0.00}$ &		           & $0.00 ^{+0.34 }_{-0.00}$  &                            &                            &                                               \\
100          & $0.00 ^{+0.47 }_{-0.00}$ &			   & $0.00 ^{+0.40 }_{-0.00}$  & $ 0.00 ^{+5.29 }_{-0.00}$  & $0.69 ^{+10.02}_{-0.69}$   & $ 7.40 ^{+ 4.65 }_{-3.80}$                    \\
105          & $0.17 ^{+0.44 }_{-0.17}$ &			   & $0.00 ^{+0.52 }_{-0.00}$  & $ 2.97 ^{+3.35 }_{-2.97}$  & $0.81 ^{ +9.51}_{-0.81}$   & $ 8.56 ^{+ 4.82 }_{-4.10}$                    \\
110          & $0.44 ^{+0.43 }_{-0.39}$ & $0.00 ^{+7.73 }_{-0.00}$ & $0.39 ^{+0.42 }_{-0.38}$  & $ 0.00 ^{+3.56 }_{-0.00}$  & $0.00 ^{ +8.07}_{-0.00}$   & $ 4.32 ^{+ 3.84 }_{-3.32}$                    \\
115          & $0.96 ^{+0.55 }_{-0.55}$ & $1.99 ^{+3.40 }_{-1.99}$ & $0.81 ^{+0.59 }_{-0.52}$  & $ 1.62 ^{+4.37 }_{-1.62}$  & $0.00 ^{ +7.38}_{-0.00}$   & $ 7.23 ^{+ 5.13 }_{-4.56}$                    \\
120          & $1.60 ^{+0.70 }_{-0.64}$ & $0.72 ^{+2.24 }_{-0.72}$ & $1.36 ^{+0.72 }_{-0.64}$  & $11.85 ^{+5.39 }_{-4.53}$  & $0.00 ^{ +7.51}_{-0.00}$   & $ 8.51 ^{+ 6.03 }_{-5.07}$                    \\
125          & $1.54 ^{+0.77 }_{-0.73}$ & $0.00 ^{+1.78 }_{-0.00}$ & $1.72 ^{+0.92 }_{-0.87}$  & $ 7.81 ^{+4.61 }_{-4.42}$  & $0.00 ^{ +8.44}_{-0.00}$   & $ 9.49 ^{+ 6.60 }_{-6.28}$                    \\
130          & $1.29 ^{+0.75 }_{-0.74}$ & $0.12 ^{+1.30 }_{-0.12}$ & $1.94 ^{+1.10 }_{-1.07}$  & $ 2.55 ^{+4.20 }_{-2.55}$  & $0.00 ^{ +9.48}_{-0.00}$   & $11.63 ^{+ 8.04 }_{-6.82}$                    \\
135          & $0.77 ^{+0.66 }_{-0.64}$ & $0.00 ^{+1.09 }_{-0.00}$ & $2.24 ^{+1.45 }_{-1.33}$  & $ 0.69 ^{+6.37 }_{-0.69}$  & $0.00 ^{+11.89}_{-0.00}$   & $10.55 ^{+ 8.71 }_{-7.15}$                    \\
140          & $0.83 ^{+0.64 }_{-0.65}$ & $0.00 ^{+0.49 }_{-0.00}$ & $2.42 ^{+1.76 }_{-1.80}$  & $ 5.15 ^{+5.24 }_{-4.91}$  & $4.15 ^{+14.11}_{-4.15}$   & $12.80 ^{+ 9.63 }_{-8.31}$                    \\
145          & $0.00 ^{+0.67 }_{-0.00}$ & $0.00 ^{+0.63 }_{-0.00}$ & $1.26 ^{+2.72 }_{-1.26}$  & $ 6.56 ^{+6.01 }_{-5.72}$  & $5.89 ^{+19.57}_{-5.89}$   & $15.81 ^{+10.70 }_{-9.29}$                    \\
150          & $0.00 ^{+0.48 }_{-0.00}$ & $0.00 ^{+0.50 }_{-0.00}$ & $3.68 ^{+3.47 }_{-3.54}$  & $ 0.00 ^{+7.21 }_{-0.00}$  & $7.90 ^{+29.12}_{-7.90}$   & $14.07 ^{+10.71 }_{-9.86}$                    \\
155          & $0.00 ^{+0.36 }_{-0.00}$ & $0.00 ^{+0.37 }_{-0.00}$ &                           &			    &		                 &                                               \\
160          & $0.00 ^{+0.30 }_{-0.00}$ & $0.00 ^{+0.29 }_{-0.00}$ &                           &			    &		                 &                                               \\
165          & $0.00 ^{+0.25 }_{-0.00}$ & $0.00 ^{+0.25 }_{-0.00}$ &			       &		            &		 	         &                                               \\
170          & $0.00 ^{+0.46 }_{-0.00}$ & $0.00 ^{+0.45 }_{-0.00}$ &			       &		            &		 	         &                                               \\
175          & $0.29 ^{+0.44 }_{-0.29}$ & $0.30 ^{+0.42 }_{-0.30}$ &			       &		            &		 	         &                                               \\
180          & $0.28 ^{+0.52 }_{-0.28}$ & $0.32 ^{+0.52 }_{-0.32}$ &			       &			    &		                 &                                               \\
185          & $0.80 ^{+0.70 }_{-0.69}$ & $0.99 ^{+0.76 }_{-0.70}$ &			       &		            &			         &                                               \\
190          & $1.31 ^{+1.00 }_{-0.85}$ & $1.55 ^{+1.00 }_{-0.99}$ &			       &		            &			         &                                               \\
195          & $1.91 ^{+1.14 }_{-1.11}$ & $2.32 ^{+1.28 }_{-1.15}$ &			       &			    &		                 &                                               \\
200          & $2.30 ^{+1.42 }_{-1.19}$ & $3.02 ^{+1.55 }_{-1.38}$ &			       &		            &			         &                                               \\
 \end{tabular}
\end{ruledtabular}
 \end{center}
 \end{table*}

\section{Fermiophobic model interpretation}
\label{sec:fp}

A number of theoretical models incorporate a Higgs boson with couplings 
to massive bosons as predicted by the SM, but negligible or zero couplings 
to fermions~\cite{fermiophobic1,fermiophobic2,fermiophobic3,fermiophobic4}.
We denote these as fermiophobic Higgs models (FHM).  Within these models 
$gg\rightarrow H$ production is negligible, as this mechanism is mediated 
at lowest order by quark loops and only higher-order weak interactions 
involving $W$ and $Z$ bosons contribute for FHM~\cite{anastasiou}.  Within 
the FHM interpretation, production rates for {\it WH}, {\it ZH}, and VBF are assumed 
to be as predicted by the SM, while the production rate for $t{\bar{t}}H$ 
is assumed to be negligible.  Higgs boson decay branching ratios to pairs 
of fermions and pairs of gluons are also set to zero.  In addition, the 
decay width $\Gamma(H\rightarrow\gamma\gamma)$ is enhanced since quark-loop
contributions, which subtract from the larger $W$-loop contribution, are 
absent.  The complete set of decay branching ratios assumed within the FHM
interpretation are listed in Table~\ref{tab:fpbr}.

\begin{figure}[htb] \begin{centering}
\includegraphics[width=0.85\columnwidth]{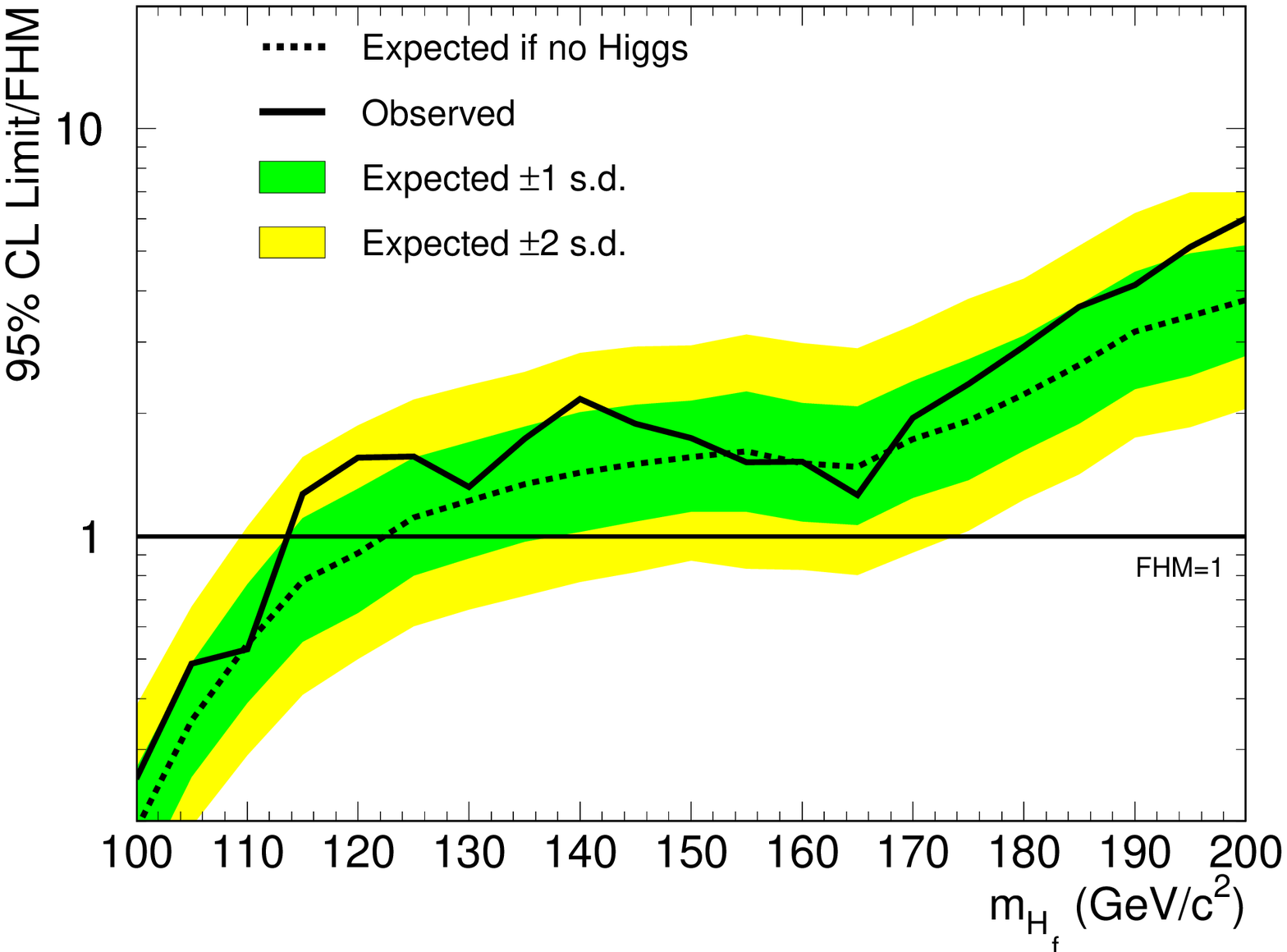}
\caption{
\label{fig:fplimits} 
Observed and expected (median, for the background-only hypothesis)
95\% C.L. upper limits on Higgs boson production within the FHM
interpretation as a function of Higgs boson mass.  The limits are
expressed as a multiple of the expected rate in the FHM for
hypothesized test masses in 5~GeV/$c^2$ increments between 100 and
200~GeV/$c^2$.  The individual points are joined together by straight
lines for better readability. The shaded bands indicate the 68\% and
95\% probability regions in which the limits are expected to fluctuate
in the absence of signal.}
\end{centering}
\end{figure}

\begin{table*}
\caption{\label{tab:fpbr}  
Decay branching fractions of the Higgs boson in FHM computed with {\sc
hdecay}~\protect{\cite{hdecay}}.  Also listed are the observed 95\%
credibility level upper limits on the signal rate relative to FHM
expectations, and the median expected limits assuming no signal is
present.}
\vspace{0.3cm}
\vspace{0.3cm}
\begin{ruledtabular}
\begin{tabular}{lccccc} 
$m_H$ (GeV/$c^2$) & $Br(\gamma\gamma)$ & $Br(W^+W^-)$ & $Br(ZZ)$    &   $R_{95}^{\rm{FHM}}$ & $R_{95,{\rm{exp}}}^{\rm{FHM}}$  \\ \hline    
 100  &   0.185   & 0.735  &     0.0762 &     0.25  &     0.19\\
 105  &   0.104   & 0.816  &     0.0733 &     0.49  &     0.35\\
 110  &   0.0603  & 0.853  &     0.0788 &     0.53  &     0.54\\
 115  &   0.0366  & 0.866  &     0.0887 &     1.27  &     0.78\\
 120  &   0.0233  & 0.869  &     0.0993 &     1.56  &     0.91\\
 125  &   0.0156  & 0.868  &     0.109     &     1.57  &     1.11\\
 130  &   0.0107  & 0.867  &     0.116     &     1.32  &     1.22\\
 135  &   $7.59\times 10^{-3}$  & 0.866  &     0.120     &     1.74  &     1.34\\
 140  &   $5.44\times 10^{-3}$  & 0.868  &     0.121     &     2.17  &     1.43\\
 145  &   $3.90\times 10^{-3}$  & 0.874  &     0.118     &     1.89  &     1.51\\
 150  &   $2.73\times 10^{-3}$  & 0.886  &     0.108     &     2.33  &     1.57\\
 155  &   $1.76\times 10^{-3}$  & 0.909  &     0.0871 &     1.52  &     1.62\\
 160  &   $8.35\times 10^{-4}$  & 0.951  &     0.0466 &     1.53  &     1.51\\
 165  &   $3.34\times 10^{-4}$  & 0.975  &     0.0236 &     1.26  &     1.48\\
 170  &   $2.26\times 10^{-4}$  & 0.975  &     0.0246 &     1.95  &     1.73\\
 175  &   $1.79\times 10^{-4}$  & 0.966  &     0.0332 &     2.36  &     1.92\\
 180  &   $1.48\times 10^{-4}$  & 0.939  &     0.0609 &     2.92  &     2.23\\
 185  &   $1.18\times 10^{-4}$  & 0.848  &     0.152     &     3.66  &     2.63\\
 190  &   $9.79\times 10^{-5}$  & 0.788  &     0.212     &     4.13  &     3.17\\
 195  &   $8.52\times 10^{-5}$  & 0.759  &     0.241     &     5.11  &     3.47\\
 200  &   $7.59\times 10^{-5}$  & 0.742  &     0.258     &     6.02  &     3.80\\
\end{tabular}
\end{ruledtabular}
\end{table*}

Previous searches for a fermiophobic Higgs boson at the Tevatron excluded 
signals with masses smaller than 119~GeV/$c^2$~\cite{prevtevfhm,prevcdffhm,prevd0fhm}; the expected 
exclusion was also $m_{H_f}<119$~GeV/$c^2$.  The ATLAS and CMS Collaborations 
excluded $m_{H_f}$ in the ranges 110.0--118.0~GeV/$c^2$ and 119.5--121.0~GeV/$c^2$ 
using diphoton final states~\cite{atlasfhm} and in the range 110--194~GeV/$c^2$ by 
combining multiple final states~\cite{cmsfhm}.

Dedicated searches are conducted for $H\rightarrow\gamma\gamma$ within
the FHM interpretation to optimize the sensitivity for the different
event kinematic properties associated with the dominant Higgs boson
production mechanisms.  FHM Higgs bosons are produced in association
with vector bosons, or recoiling from jets in the case of VBF. As a
result, the Higgs boson $p_T$ spectrum is shifted to higher values for
the FHM than the SM, where the dominant production mechanism is
$gg\rightarrow H$.  Potential signal contributions from {\it WH}, {\it
ZH}, and VBF production included in the SM $H\rightarrow W^+W^-$ and
$H\rightarrow ZZ$ search channels are also incorporated.  In the
$H\rightarrow W^+W^-$ search sub-channel focusing on events with
opposite-charge leptons and two or more reconstructed jets, where
potential signal contributions from these production mechanisms are
significant, the final discriminant used for the FHM interpretation
has been re-optimized to focus on the expected event kinematic
properties of the relevant signal processes.  The FHM search is
performed over the range $100\leq m_H\leq 200$~GeV/$c^2$.

No evidence for a fermiophobic Higgs boson is found in the data, and
upper limits are set on the production rate relative to the FHM
expectation.  These limits are shown in Fig.~\ref{fig:fplimits} and
listed in Table~\ref{tab:fpbr}.  We exclude a fermiophobic Higgs boson
in the mass range $100<m_H<113$~GeV/$c^2$, and expect to exclude
$100<m_H<122$~GeV/$c^2$ in the absence of a Higgs boson signal.

\section{Fourth-generation model interpretation and model-independent limit on $gg\rightarrow H$ production}
\label{sec:sm4}

The lowest-order process mediating the {\it ggH} coupling in the SM is a
quark triangle-loop, with the dominant contribution coming from the
top quark, and a smaller contribution from the bottom quark.  The
model tested here is the standard model with a fourth sequential
generation of fermions (SM4).  The masses of the components of the
fourth generation are assumed to be larger than the mass bounds from
collider experiments.  In the SM4, the up-type ($u_4$) and down-type
($d_4$) quarks would contribute approximately with the same magnitude
as the top quark to the {\it ggH} coupling, resulting in approximately a
factor of nine increase in the $gg\rightarrow H$ production cross
section and the $H\rightarrow gg$ decay width~\cite{sm41,sm42,sm43}.
The enhancement is modified by resonant structure in the quark loop
(the top quark contributes most strongly when $m_H\approx 2m_t$),
electroweak contributions~\cite{sm42}, and QCD radiative
corrections~\cite{sm43}.  The decay branching fractions of the Higgs
boson may further be modified by the presence of a fourth neutrino
($\nu_4$), which may have been too heavy to be discovered at LEP, or
due to decays to a heavy fourth-generation charged lepton $\ell_4$.
We do not include acceptance for these decays in our predicted signal yields.  The
precision electroweak constraints that place an upper bound on the SM
Higgs boson mass~\cite{sm41} are significantly relaxed in the SM4, allowing Higgs boson
masses up to 750~GeV/$c^2$.

 The production cross section for $gg\rightarrow H$ is computed in
Ref.~\cite{sm43} for two scenarios of $m_{u_4}$ and $m_{d_4}$, but the
production rates do not depend significantly on these masses, once
they are large enough to evade experimental bounds.  If $2m_{\ell
_4}<m_H$ and $2m_{\nu _4}<m_H$, the decay branching ratios have a
large impact on our ability to test the model.  In both scenarios we
assume $m_{u_4}=450$~GeV/$c^2$ and $m_{d_4}=400$~GeV/$c^2$.  In the
first scenario, called the {\it high-mass} scenario, we assume
$m_{\ell _4}=m_{\nu _4}=1000$~GeV/$c^2$, and in the second scenario,
the {\it low-mass} scenario, $m_{\ell _4}=100$~GeV $c^2$ and $m_{\nu
_4}=80$~GeV/$c^2$.

We search for $gg\rightarrow H$ production primarily in the
$H\rightarrow W^+W^-$ decay mode, but the $H\rightarrow ZZ$ decay mode
also contributes, particularly for $m_H>200$~\gevcc.  The
$H\rightarrow\gamma\gamma$ channels contribute in the SM mainly
through the $gg\rightarrow H$ production, but this decay mode is
suppressed due to negative contributions of the quark loops relative
to the $W$-mediated loop in $H\rightarrow\gamma\gamma$ decay.  We
therefore include only the $H\rightarrow W^+W^-$ and $H\rightarrow
ZZ\rightarrow\ell^+\ell^-\ell^+\ell^-$ searches in this
interpretation.

Previous interpretations of SM Higgs boson searches within the context of a 
fourth generation of fermions at Tevatron excluded $131<m_H<207$~GeV/$c^2$~\cite{PRDRC}. 
Searches with similar sensitivity were performed by the ATLAS~\cite{atlas4g} and CMS~\cite{cms4g} 
Collaborations, excluding $140<m_H<185$~GeV/$c^2$ and $144<m_H<207$~GeV/$c^2$, 
respectively.  A more recent search by the CMS Collaboration excluded the mass 
range $110<m_H<600$~GeV/$c^2$~\cite{cms4g-new}.

\begin{figure}[htb] \begin{centering}
\includegraphics[width=0.85\columnwidth]{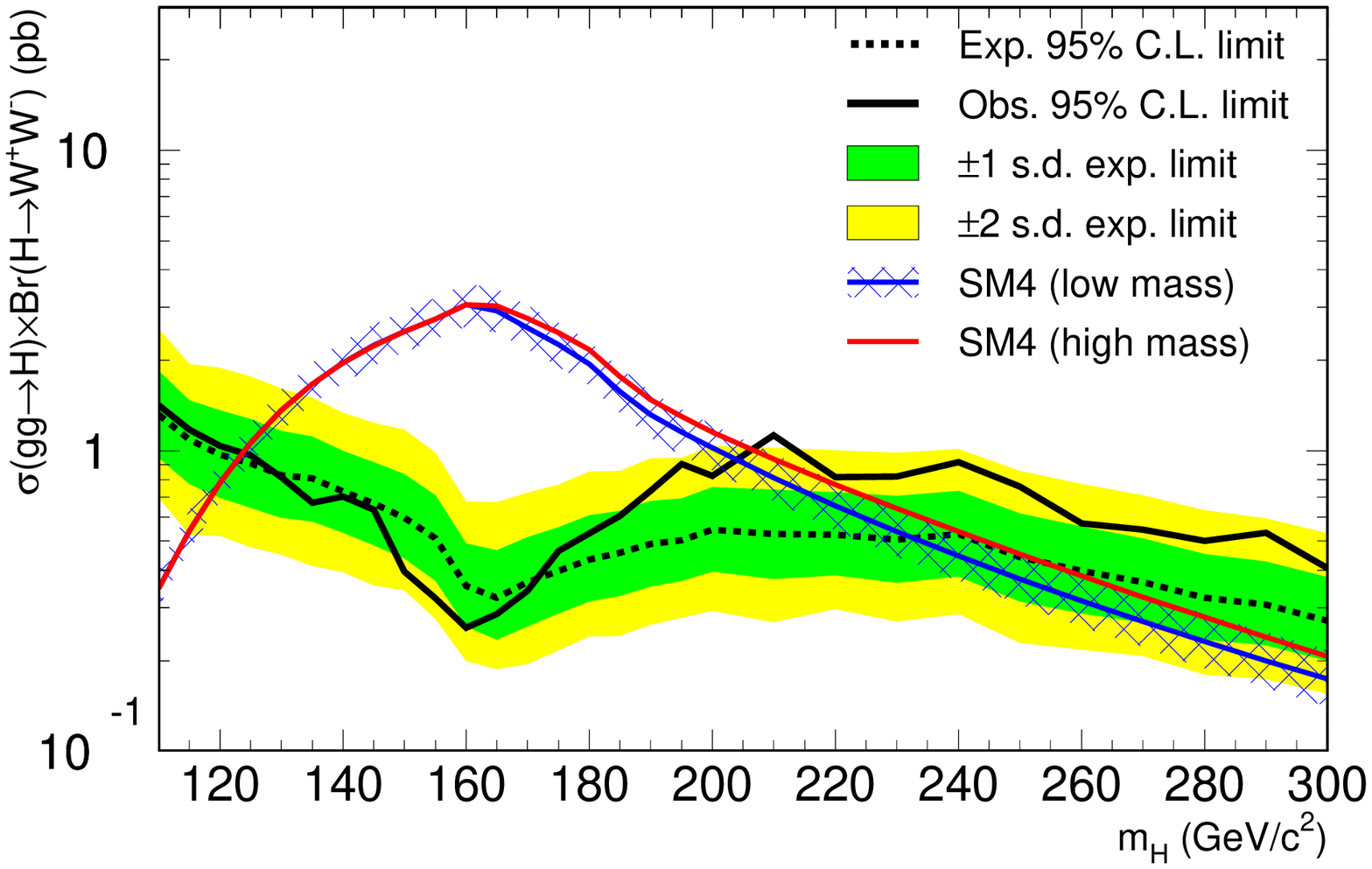}
\caption{
\label{fig:gghlimits} 
Observed and expected (median, for the background-only hypothesis)
95\% C.L. upper limits on the production rate for $gg\rightarrow
H\rightarrow W^+W^-$ in picobarns, as functions of the Higgs boson mass.  The
points are joined by straight lines for better readability. The bands
indicate the 68\% and 95\% probability regions where the limits can
fluctuate, in the absence of signal.  Also shown are the predictions
for the two SM4 scenarios described in the text.}
\end{centering}
\end{figure}

The first step is to set a limit on
$\sigma(gg\rightarrow H)\times Br(H\rightarrow W^+W^-)$, which can be
interpreted in a variety of models.  We assume the SM value for the
ratio of $Br(H\rightarrow ZZ)/Br(H\rightarrow W^+W^-)$ when combining
the $ZZ$ results, an assumption which is accurate in
the SM4.  The $H\rightarrow W^+W^-$ channels are re-optimized for this
search by training the discriminants to separate only the
$gg\rightarrow H$ mode from the background, ignoring the {\it WH}, {\it ZH},
and VBF production modes.  In setting upper limits on the
$gg\rightarrow H$ production cross section, we also ignore the
acceptance for {\it WH}, {\it ZH}, and VBF production, which yields
conservative limits.  In setting limits on $\sigma(gg\rightarrow
H)\times Br(H\rightarrow W^+W^-)$, we do not include uncertainties in
theoretical predictions of the production cross section or the decay
branching ratio, but we include the theoretical uncertainties on
the relative signal expectations in the 0-jet, 1-jet, and 2+jet
event selections in the $H\rightarrow W^+W^-$ searches.  We search for
Higgs bosons in the mass range $110<m_H<300$~GeV/$c^2$, in which the
analysis is expected to be sensitive to the SM4.  Limits on $\sigma(gg\rightarrow H)\times Br(H\rightarrow
W^+W^-)$ are listed in Table~\ref{tab:4glimits}, and are shown in
Fig.~\ref{fig:gghlimits}.

\begin{table*}
\caption{\label{tab:4glimits}
Observed and median expected upper limits on $\sigma\times
Br(H\rightarrow W^+W^-)$ at the 95\% C.L., as well as the
predicted $gg\rightarrow H$ production cross sections and decay
branching fractions in the SM4 with $m_{\nu_4}=80$~\gevcc,
$m_{\ell_4}=100$~\gevcc, $m_{d_4}=400$~\gevcc, and
$m_{u_4}=450$~\gevcc.}
\vspace{0.3cm}
\begin{ruledtabular}
\begin{tabular}{lccccccccc}
     & Obs & Exp & &\multicolumn{4}{|c}{Low-mass scenario}&\multicolumn{2}{|c}{High-mass scenario} \\
$m_H$ (GeV/$c^2$) 
     & limit (pb)&limit (pb)&$\sigma(gg\rightarrow H)$~(pb)
                                      &\multicolumn{1}{|c}{$Br(W^+W^-)$}
                                                   & $Br(ZZ)$                & $Br(\nu_4{\bar{\nu}}_4)$ & $Br(\ell_4^+\ell_4^-)$ 
                                                                                                         & \multicolumn{1}{|c}{$Br(W^+W^-)$}   & $Br(ZZ)$      \\ \hline
 110 &     1.42  &     1.32 & 12.3    & 0.0283     & $2.62\times 10^{-3}$    &   0.00      &   0.00      & 0.0283     & $2.62\times 10^{-3}$  \\	
 115 &     1.18  &     1.09 & 10.7    & 0.0505     & $5.17\times 10^{-3}$    &   0.00      &   0.00      & 0.0505     & $5.17\times 10^{-3}$  \\	
 120 &     1.04  &     0.97 & 9.38    & 0.0834     & $9.52\times 10^{-3}$    &   0.00      &   0.00      & 0.0834     & $9.52\times 10^{-3}$  \\	
 125 &     0.97  &     0.91 & 8.24    & 0.129      & 0.0161                  &    0.00     &   0.00      & 0.129      & 0.0161  \\	
 130 &     0.81  &     0.83 & 7.26    & 0.188      & 0.0251                  &    0.00     &   0.00      & 0.188      & 0.0251  \\	
 135 &     0.67  &     0.81 & 6.41    & 0.260      & 0.0362                  &    0.00     &   0.00      & 0.260      & 0.0362  \\	
 140 &     0.70  &     0.73 & 5.68    & 0.346      & 0.0483                  &    0.00     &   0.00      & 0.346      & 0.0483  \\	
 145 &     0.63  &     0.67 & 5.05    & 0.443      & 0.0597                  &    0.00     &   0.00      & 0.443      & 0.0597  \\	
 150 &     0.40  &     0.60 & 4.50    & 0.553      & 0.0672                  &    0.00     &   0.00      & 0.553      & 0.0672  \\	
 155 &     0.32  &     0.51 & 4.02    & 0.681      & 0.0653                  &    0.00     &   0.00      & 0.681      & 0.0653  \\	
 160 &     0.26  &     0.35 & 3.60    & 0.850      & 0.0409                  &    0.00     &   0.00      & 0.850      & 0.0409  \\	
 165 &     0.29  &     0.32 & 3.22    & 0.906      & 0.0199                  &   0.0387    &   0.00      & 0.942      & 0.0207  \\	
 170 &     0.34  &     0.36 & 2.89    & 0.888      & 0.0207                  &   0.0672    &   0.00      & 0.952      & 0.0222  \\	
 175 &     0.46  &     0.40 & 2.60    & 0.863      & 0.0279                  &   0.0893    &   0.00      & 0.948      & 0.0306  \\	
 180 &     0.53  &     0.43 & 2.35    & 0.828      & 0.0510                  &   0.104     &   0.00      & 0.925      & 0.0569  \\	
 185 &     0.61  &     0.46 & 2.12    & 0.742      & 0.138                   &   0.107     &   0.00      & 0.831      & 0.154   \\	
 190 &     0.73  &     0.49 & 1.92    & 0.687      & 0.194                   &   0.109     &   0.00      & 0.770      & 0.217   \\	
 195 &     0.90  &     0.50 & 1.74    & 0.661      & 0.217                   &   0.112     &   0.00      & 0.745      & 0.244   \\	
 200 &     0.83  &     0.55 & 1.58    & 0.647      & 0.230                   &   0.114     &   0.00      & 0.730      & 0.260   \\	
 210 &     1.13  &     0.53 & 1.31    & 0.620      & 0.239                   &   0.115     &  0.0187 	  & 0.715      & 0.276   \\	
 220 &     0.82  &     0.52 & 1.09    & 0.600      & 0.242                   &   0.112     &  0.0393 	  & 0.708      & 0.284   \\	
 230 &     0.82  &     0.50 & 0.912   & 0.588      & 0.242                   &   0.108     &  0.0551 	  & 0.703      & 0.290   \\	
 240 &     0.92  &     0.53 & 0.767   & 0.581      & 0.244                   &   0.104     &  0.0663 	  & 0.700      & 0.294   \\	
 250 &     0.76  &     0.44 & 0.649   & 0.577      & 0.245                   &   0.0991    &  0.0738 	  & 0.697      & 0.296   \\	
 260 &     0.57  &     0.40 & 0.551   & 0.575      & 0.247                   &   0.0944    &  0.0787 	  & 0.695      & 0.299   \\	
 270 &     0.54  &     0.37 & 0.470   & 0.575      & 0.250                   &   0.0898    &  0.0814 	  & 0.693      & 0.301   \\	
 280 &     0.50  &     0.32 & 0.403   & 0.576      & 0.252                   &   0.0853    &  0.0827 	  & 0.692      & 0.303   \\	
 290 &     0.53  &     0.31 & 0.347   & 0.577      & 0.255                   &   0.0810    &  0.0829 	  & 0.690      & 0.305   \\	
 300 &     0.41  &     0.27 & 0.300   & 0.579      & 0.258                   &   0.0770    &  0.0823  	  & 0.689      & 0.306   \\
\end{tabular}
\end{ruledtabular}
\end{table*}

\begin{figure}[htb] \begin{centering}
\includegraphics[width=0.85\columnwidth]{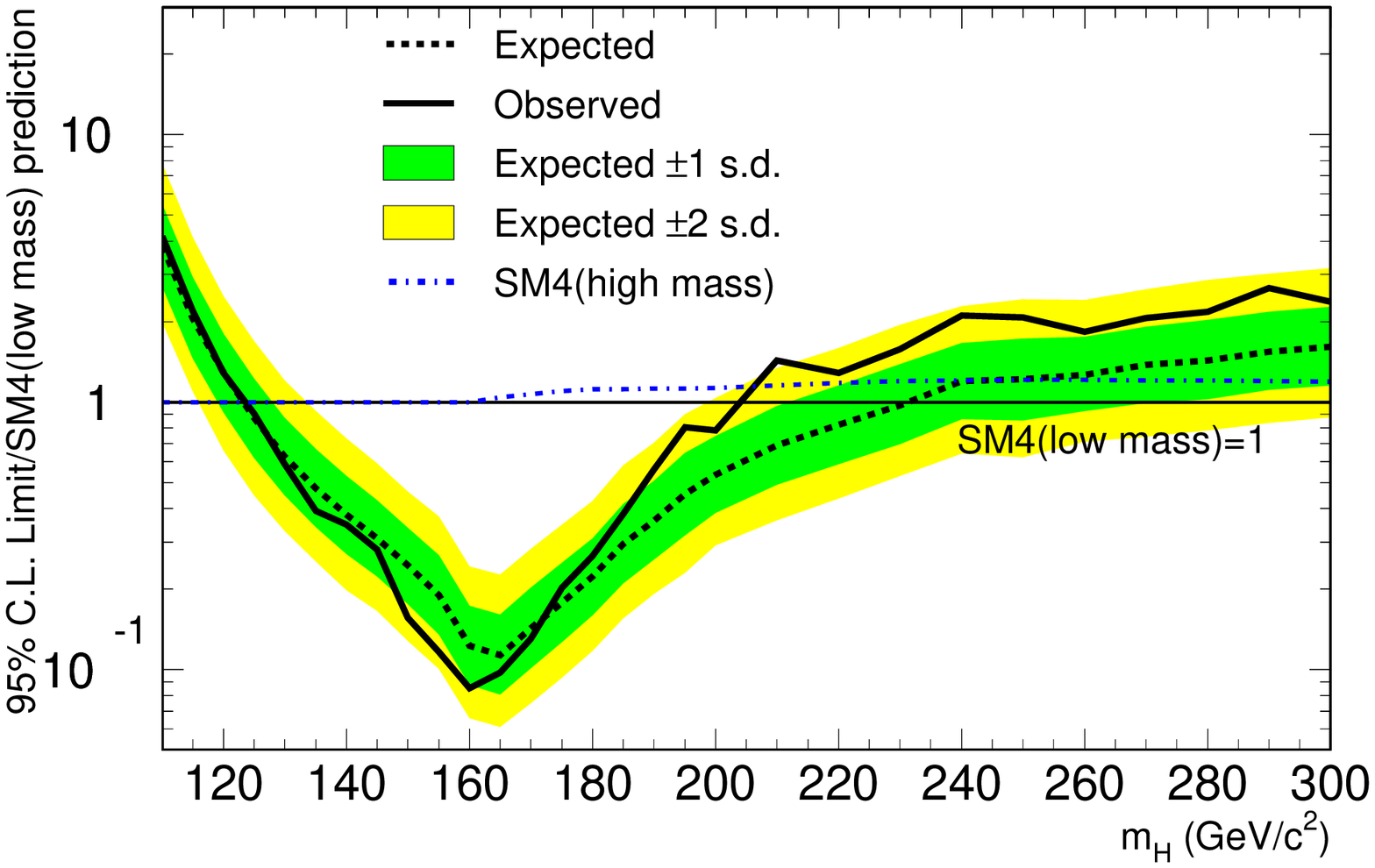}
\caption{
\label{fig:sm4limits} 
Observed and expected (median, for the background-only hypothesis)
95\% C.L. upper limits on Higgs boson production has a function of
Higgs boson mass, in the SM4 model in the low-mass scenario, which
gives the loosest mass bounds.  The prediction for the high-mass
scenario is also shown.  The limits are expressed as a multiple of the
SM4 prediction.  The points are joined by straight lines for better
readability. The bands indicate the 68\% and 95\% probability regions
where the limits can fluctuate, in the absence of signal. }
\end{centering}
\end{figure}

The second step in the SM4 interpretation is to consider specific
model scenarios. In this step we reintroduce the theoretical
uncertainties on the predicted cross sections due to QCD factorization
and renormalization scale and PDF uncertainties.  The limits obtained
are shown in Fig.~\ref{fig:sm4limits} as multiples of the predictions
in the two scenarios.  In the low-mass scenario, we exclude the range
$124<m_H<203$~GeV/$c^2$ at the 95\% C.L., and expect to exclude
$123<m_H<231$~GeV/$c^2$.  In the high-mass scenario, the lack of
fourth-generation leptonic and neutrino decays provides more expected
signal in the remaining visible decays. We exclude the range
$124<m_H<206$~GeV/$c^2$ at the 95\% C.L., and expect to exclude the
range $123<m_H<245$~GeV/$c^2$.

\section{Constraints on fermionic and bosonic couplings}
\label{sec:couplings}

Following the recent LHC observations of a new Higgs-like particle
with a mass of approximately 125 GeV/$c^2$, we focus on this mass
hypothesis and test the couplings of the new particle, assuming that the
mild observed excesses in CDF's Higgs boson searches originate from this
source.  Similar studies of the couplings have been performed by
CMS~\cite{cmscoupling} and ATLAS~\cite{atlascoupling}.

We assume that the production and decay of the Higgs-like particle
follows the predictions of the SM Higgs boson, but with modified
coupling strengths to fermions, the $W$ boson, and the $Z$ boson.  We
follow the procedures and notation of Ref.~\cite{lhcxscoupling}, and
scale all Higgs boson couplings to fermions, regardless of flavor, by
the factor $\kappa_f$; we scale the $HWW$ coupling by the factor
$\kappa_W$, and the $HZZ$ coupling by the factor $\kappa_Z$.  The
predicted signal rates in each production and decay mode are functions
of the SM predictions and the factors $\kappa_f$, $\kappa_W$, and
$\kappa_Z$.  The SM predictions are obtained by setting
$\kappa_f=\kappa_W=\kappa_Z=1$.  Because the $\kappa$ factors scale
the couplings, the production rates and decay widths are quadratic
functions of the coupling scale factors.  The decay branching ratios
are computed from the decay widths and thus are ratios of quadratic
functions of the coupling scale factors.

For each of the studies described below, we assume a uniform prior
probability density in one or more of the coupling scale factors and
compute the posterior probability density using all of CDF SM Higgs
boson search results, integrating over systematic uncertainties.
One-dimensional intervals are quoted as the shortest set of intervals
containing 68\% of the integral of the posterior density, and the
two-dimensional contours are those with the smallest areas containing
68\% and 95\% of the integral of the posterior density.  The values
that maximize the posterior probability are quoted as best-fit values.

We study both positive and negative values of the coupling scale
factors, although little information on the relative signs of the
couplings remains after squaring the amplitudes.  The posterior
probability densities have multiple maxima, possibly asymmetric due to
interference terms in the production and decay in some modes.  The
$H\gamma\gamma$ coupling has a destructive interference term arising
from the contributions from fermion loops and the $W$-boson loop that
introduces a term linear in $\kappa_W$ and $\kappa_f$.  This term
breaks the ambiguity of the relative sign between $\kappa_W$ and
$\kappa_f$, although the contribution from the
$H\rightarrow\gamma\gamma$ channels is weak in the analyses presented
here.  A smaller interference term exists in the {\it ggH} coupling,
in which the dominant fermion-loop contributions interfere
constructively with two-loop electroweak
contributions~\cite{anastasiou,actis2008}.  A global sign on all Higgs
boson couplings is unobservable in the current analysis.

\begin{figure} \begin{centering}
\includegraphics[width=0.95\columnwidth]{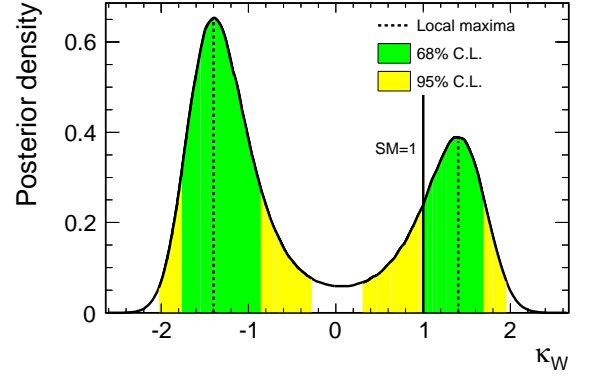}
\caption{
\label{fig:kappaw} Posterior probability distribution for $\kappa_{W}$ from the 
combination of all CDF search channels.  In performing this fit, the values 
for $\kappa_{Z}$ and $\kappa_{f}$ are fixed to their SM values ($\kappa_{Z}
= \kappa_{f} = 1$).}
\end{centering}
\end{figure}

\begin{figure} \begin{centering}
\includegraphics[width=0.95\columnwidth]{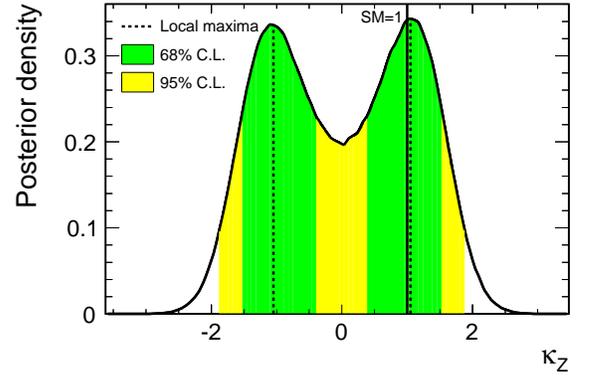}
\caption{
\label{fig:kappaz} Posterior probability distribution for $\kappa_{Z}$ from the 
combination of all CDF search channels.  In performing this fit, the values 
for $\kappa_{W}$ and $\kappa_{f}$ are fixed to their SM values ($\kappa_{W}
= \kappa_{f} = 1$).}
\end{centering}
\end{figure}

\begin{figure} \begin{centering}
\includegraphics[width=0.95\columnwidth]{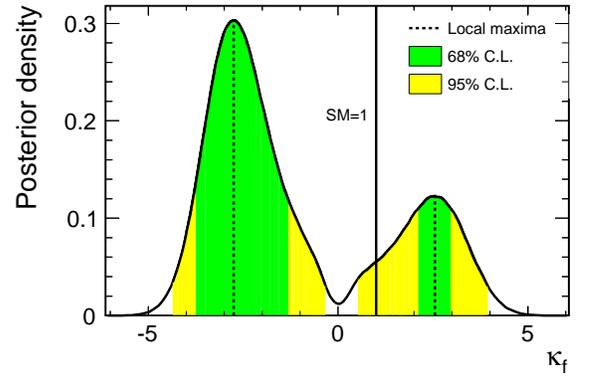}
\caption{
\label{fig:kappaf} Posterior probability distribution for $\kappa_{f}$ from the 
combination of all CDF search channels.  In performing this fit, the values 
for $\kappa_{W}$ and $\kappa_{Z}$ are fixed to their SM values ($\kappa_{W}
= \kappa_{Z} = 1$).}
\end{centering}
\end{figure}

We study each coupling scale factor independently, holding the others
fixed to their SM values, and then to study the fits by relaxing the
assumptions one at a time.  We first study $\kappa_{W}$, setting
$\kappa_{f}=\kappa_{Z}=1$.  The posterior probability distribution for
$\kappa_{W}$ is shown in Fig.~\ref{fig:kappaw}.  The factor
$\kappa_{W}$ is constrained to the intervals $-1.8 < \kappa_{W} <
-0.8$ and $1.0 < \kappa_{W} < 1.7$ at the 68\% C.L.  The best fit
value for $\kappa_{W}$ is --1.4.  We perform a similar fit for
$\kappa_{Z}$, setting $\kappa_{f}=\kappa_{W}=1$.  From the posterior
probability distribution shown in Fig.~\ref{fig:kappaz}, $\kappa_{Z}$
is constrained at the 68\% C.L. to the intervals $-1.5 < \kappa_{Z} <
-0.4$ and $0.4 < \kappa_{Z} < 1.5$.  The best fit value for
$\kappa_{Z}$ is 1.05.  We also perform a one-dimensional fit for the
Higgs boson coupling to fermions, $\kappa_{f}$, setting
$\kappa_{W}=\kappa_{Z}=1$.  The posterior probability distribution
obtained from the fit is shown in Fig.~\ref{fig:kappaf}.  In this case
$\kappa_{f}$ is restricted at the 68\% C.L. to the intervals $-3.8 <
\kappa_{f} < -1.2$ and $2.0 < \kappa_{f} < 3.0$ and has a best fit
value of --2.75.

\begin{figure} \begin{centering}
\includegraphics[width=0.95\columnwidth]{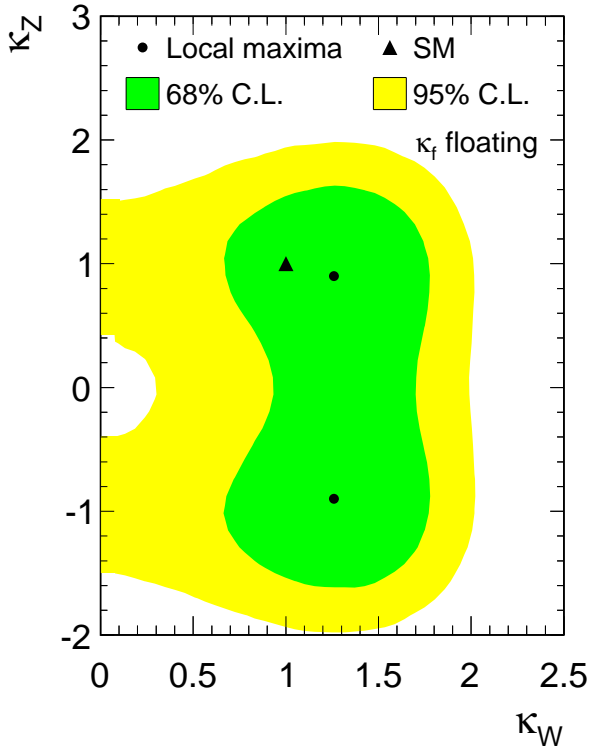}
\caption{
\label{fig:kappawz} Two-dimensional constraints in the $(\kappa_{W},\kappa_{Z})$ 
plane.  The 68\% and 95\% credibility regions are shown.  A
three-dimensional posterior probability distribution is obtained from
a fit to the observed data in all CDF search channels by floating all
three coupling parameters ($\kappa_{W}$, $\kappa_{Z}$, and
$\kappa_{f}$).  The two-dimensional constraints are obtained by
integrating the three-dimensional posterior probability density over
$\kappa_{f}$.  
}
\end{centering}
\end{figure}

We also constrain the allowed parameter space in the two-dimensional
$(\kappa_{W},\kappa_{Z})$ plane.  A fit to the observed data is
performed allowing all three coupling parameters to float.
Two-dimensional constraints on $(\kappa_{W},\kappa_{Z})$ are obtained
from the resulting three-dimensional posterior probability
distribution by integrating over $\kappa_{f}$.  The smallest regions
containing 68\% and 95\% of the integral of the posterior probability
density are shown in Fig.~\ref{fig:kappawz}.  As a result of the
global sign ambiguity in the couplings, the value of the posterior
probability at $(-\kappa_{W},\kappa_{Z})$ is equal to the value at
$(\kappa_{W},-\kappa_{Z})$.  Similarly, the value of the posterior
probability at $(-\kappa_{W},-\kappa_{Z})$ is equal to the value at
$(\kappa_{W},\kappa_{Z})$.  The posterior probability distribution in
Fig.~\ref{fig:kappawz} is displayed only for positive values of
$\kappa_{W}$.  The local maxima within the regions of positive and
negative $\kappa_{Z}$ are ($\kappa_{W}$ = 1.3, $\kappa_{Z}$ = 0.9) and
($\kappa_{W}$ = 1.3, $\kappa_{Z}$ = --0.9).

\begin{figure} \begin{centering}
\includegraphics[width=0.95\columnwidth]{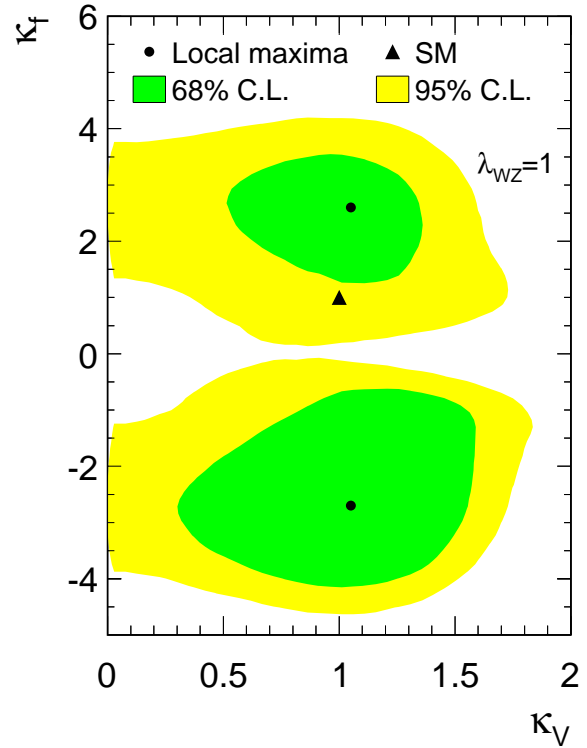}
\caption{
\label{fig:cvcf} Two-dimensional constraints in the $(\kappa_{V},\kappa_{f})$ 
plane.  The 68\% and 95\% credibility regions are shown.  A
two-dimensional posterior probability distribution is obtained from a
fit to the observed data in all CDF search channels where $\kappa_{V}$
is the generic Higgs boson coupling to $W$ and $Z$ bosons assuming the
SM value of one for $\lambda_{WZ} = \kappa_{W}/\kappa_{Z}$.}
\end{centering}
\end{figure}

Finally, we constrain the allowed parameter space within the
two-dimensional $(\kappa_{V},\kappa_{f})$ plane.  Here, $\kappa_{V}$
refers to a generic coupling of the Higgs boson to both $W$ and $Z$
bosons where the ratio $\lambda_{WZ} = \kappa_{W}/\kappa_{Z}$ is fixed
to unity.  We compute a two-dimensional posterior probability
distribution in the $(\kappa_{V},\kappa_{f})$ plane assuming a uniform
prior probability density.  The smallest regions containing 68\% and
95\% of the integral of the posterior probability density are shown in
Fig.~\ref{fig:cvcf}.  Accounting for the symmetries
$(-\kappa_{V},\kappa_{f}) = (\kappa_{V},-\kappa_{f})$ and
$(-\kappa_{V},-\kappa_{f}) = (\kappa_{V},
\kappa_{f})$ the posterior probability distribution is only displayed for positive values 
of $\kappa_{V}$.  The local maxima within the regions of positive and
negative $\kappa_{f}$ are ($\kappa_{V}$ = 1.05, $\kappa_{f}$ = 2.6)
and ($\kappa_{V}$ = 1.05, $\kappa_{f}$ = --2.7).

The results in the $(\kappa_{V},\kappa_{f})$ plane shown here are more
constraining than those previously extracted in Ref.~\cite{grojean}.
This is due to the inclusion of more channels and the use of separate
scalings for each signal component, itemized by production and decay
mode, contributing to individual search channels.  The search channels
with the most sensitivity to SM Higgs boson production measure the
product of vector boson and fermion couplings.  For example, search
modes focusing on decays to fermion pairs ($b{\bar{b}}$ and
$\tau^+\tau^-$) require production in association with a vector boson.
Conversely, searches focusing on Higgs boson decays to $W^+W^-$ and
$ZZ$ pairs obtain a majority of their sensitivity from $gg
\rightarrow H$ production, which proceeds mostly via fermionic couplings 
to the Higgs boson.  Our searches for $t{\bar{t}}H\rightarrow
t{\bar{t}}b{\bar{b}}$, on the other hand, are sensitive primarily to
$\kappa_{f}$ in both the production and decay modes.  Hence, this
search channel contributes significantly to the observed constraints
on the coupling parameters, although it provides only a small
contribution to combined CDF SM search result.  Similarly, search
channels focusing on events with same-charge di-leptons and
tri-leptons are sensitive to $\kappa_{V}$ in both the production ({\it
VH}) and decay ($H\rightarrow W^+W^-$) modes.  These channels provide
a loose constraint on $\kappa_{V}$ independent of $\kappa_{f}$ and in
the process help eliminate tails in the posterior probability
distributions.

\section{Summary}
\label{sec:summary}

In summary, we present a final combination of the CDF Higgs boson
searches.  In the context of the standard model, we exclude at the
95\% C.L. Higgs bosons with masses in the ranges
$90<m_H<102$~GeV/$c^2$ and $149<m_H<172$~GeV/$c^2$.  In the absence of
a signal, we expect to exclude the ranges $90<m_H<94$~GeV/$c^2$,
$96<m_H<106$~GeV/$c^2$, and $153<m_H<175$~GeV/$c^2$.  An excess of
data with respect to background predictions is observed, which has a
local significance of 2.0 standard deviations at $m_H=125$~GeV/$c^2$.
We exclude fermiophobic Higgs boson bosons with mass in the range
$100<m_H<113$~GeV/$c^2$, and expect to exclude $100<m_H<122$~GeV/$c^2$
in the absence of a Higgs boson signal.  In the fourth-generation
scenario incorporating the lowest possible fourth generation lepton
and neutrino masses, we exclude the range $124<m_H<203$~GeV/$c^2$ at
the 95\% C.L., and expect to exclude $123<m_H<231$~GeV/$c^2$.  The
constraints placed on the fermionic and bosonic couplings are
consistent with standard model expectations.

\begin{center}
{\bf Acknowledgments}
\end{center}

We would like to thank the authors of the {\sc hawk} program for
adapting it to the Tevatron.  We thank the Fermilab staff and the
technical staffs of the participating institutions for their vital
contributions. This work was supported by the U.S. Depar tment of
Energy and National Science Foundation; the Italian Istituto Nazionale
di Fisica Nucleare; the Ministry of Education, Culture, Sports,
Science and Technology of Japan; the Natural Sciences and Engineering
Research Council of Canada; the National Science Council of the
Republic of China; the Swiss National Science Foundation; the
A.P. Sloan Foundation; the Bundesministerium f\"ur Bildung und
Forschung, Germany; the Korean World Class University Program, the
National Research Foundation of Korea; the Science and Technology
Facilities Council and the Royal Society, UK; the Russian Foundation
for Basic Research; the Ministerio de Ciencia e Innovaci\'{o}n, and
Programa Consolider-Ingenio 2010, Spain; the Slovak R\&D Agency; the
Academy of Finland; the Australian Research Council (ARC); and the EU
community Marie Curie Fellowship contract 302103.



\newpage

\begin{table*}
\begin{center}
{\Large{\bf Auxiliary material}}
\end{center}
\end{table*}

\setcounter{figure}{0}
\setcounter{table}{0}

\begin{figure*} \begin{centering}
\includegraphics[width=0.85\textwidth]{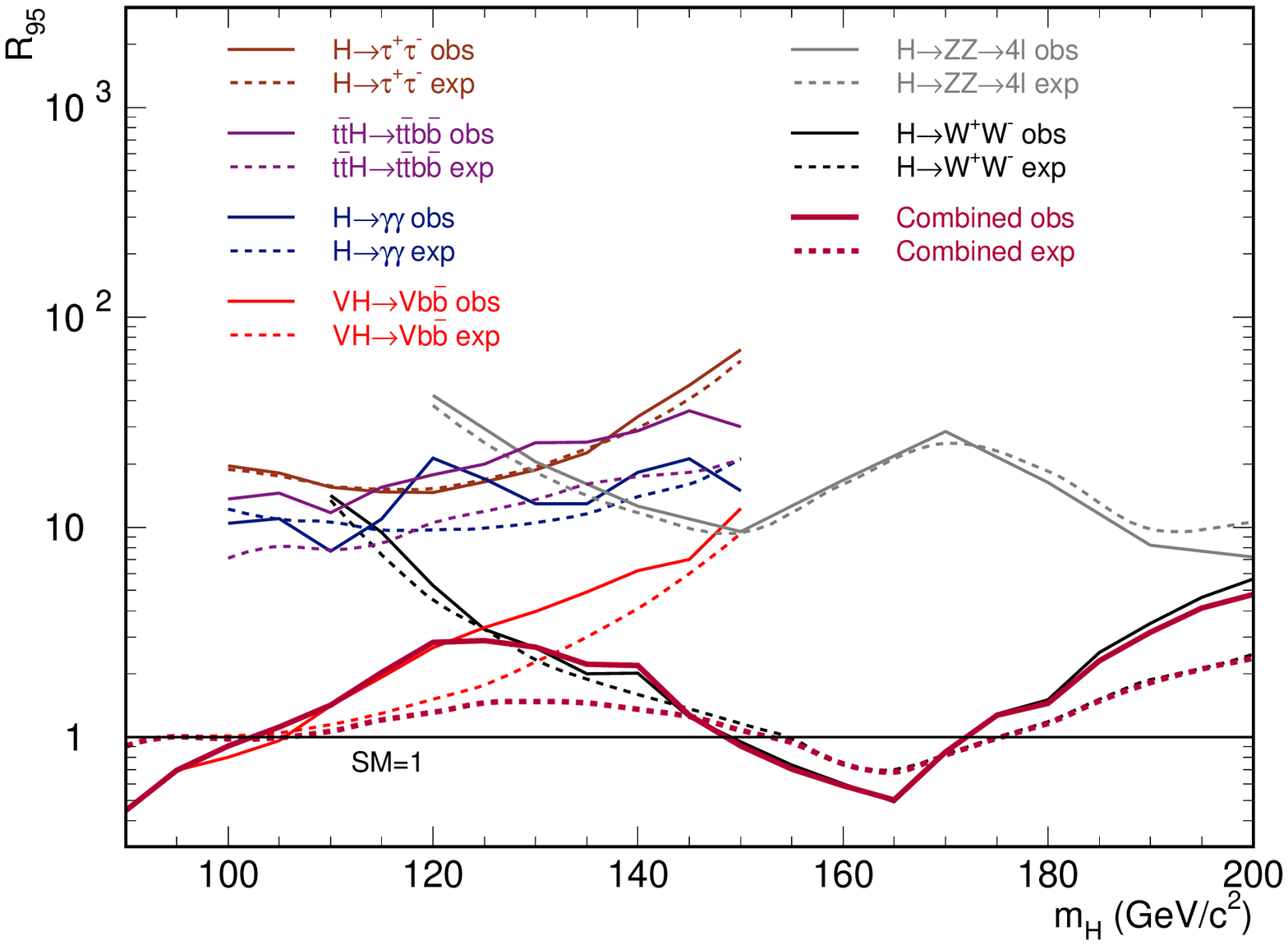}
\caption{
\label{fig:spaghetti2} 
Median expected (assuming the background-only hypothesis) and observed 
95\% C.L. upper limits on Higgs boson production relative to the SM 
expectation for combinations of search channels within each Higgs boson 
decay mode and the combination of all search channels as a function of 
Higgs boson mass in the range between 90 and 200~\gevcc.}
\end{centering}
\end{figure*}

\begin{figure*} \begin{centering}
\includegraphics[width=0.85\textwidth]{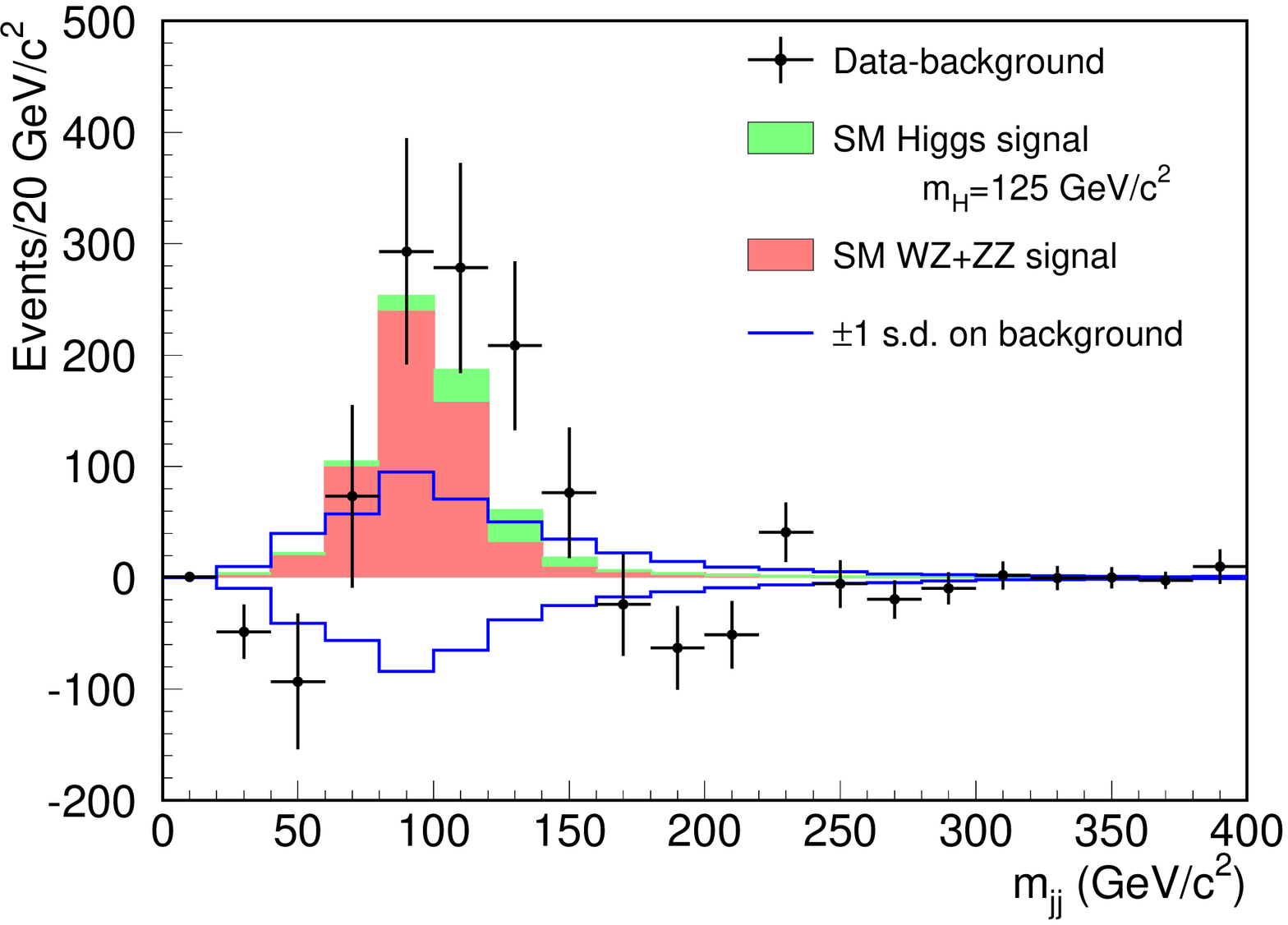}
\caption{
\label{fig:dibo2} Background-subtracted dijet invariant mass distribution 
from the combination of all search channels contributing to the $VZ$ cross 
section measurement.  Expected signal contributions from $VZ$ production 
(red) and a $m_H =$~125~GeV/$c^2$ SM Higgs boson (green) are indicated 
with the filled histograms.  The normalization of the subtracted background 
contribution, with uncertainties indicated by the unfilled histogram, is 
obtained from a fit to the data.}
\end{centering}
\end{figure*}


\begin{thebibliography}{91}
\bibitem[{not()}]{notation}
\bibinfo{note}{Here, and throughout the paper, charge conjugate states are
  implied. The label $\ell$ indicates an electron or a muon, but not a tau
  lepton. Flavor indices on neutrinos are also implied. $W^-\rightarrow
  \ell^-{\bar{\nu}}_\ell$ is written $W\rightarrow\ell\nu$.}
\bibitem[{coo()}]{coord}
\bibinfo{note}{We use a cylindrical coordinate system with origin in the center
  of the detector where $\theta$ and $\phi$ are the polar and azimuthal angles,
  respectively, and pseudorapidity is $\eta= - \rm ln$ $\rm tan(\theta/2)$. The
  missing $E_T$ (\METVEC) is defined by \METVEC$ = -\sum_iE_T^i{\hat n}_i,
  i={\rm calorimeter~tower~number}$, where ${\hat n}_i$ is a unit vector
  perpendicular to the beam axis and pointing at the $i$th calorimeter tower.
  The calculation for~\METVEC{} includes jet-energy correction and corrections
  for high-energy muons. We define \MET=$|$\METVEC$|$. The transverse momentum
  $p_T$ is defined to be $p\sin\theta$.}
\end{thebibliography}
\end{document}